\documentclass[pra,twocolumn,superscriptaddress]{revtex4}

\usepackage{graphics}
\usepackage{graphicx}
\usepackage{bm}
\usepackage{latexsym}
\usepackage{color}
\usepackage{mathrsfs}
\usepackage{braket}
\usepackage{enumerate}
\usepackage[usenames,dvipsnames]{xcolor}
\usepackage[colorlinks=true,citecolor=blue,linkcolor=blue,urlcolor=blue,backref=true]{hyperref}
\usepackage{float}
\usepackage[raggedright,tight]{subfigure}  
\usepackage{amsmath}
\usepackage{amsfonts}
\usepackage{braket}
\usepackage[normalem]{ulem}
\usepackage{cancel}
\usepackage{mathtools}
\usepackage{appendix}
\newcommand{\mb}{\mathbf}

\newcommand{\rme}{\mathrm{e}}

\newcommand{\x}{\left(\mb{x}\right)}
\newcommand{\y}{\left(\mb{y}\right)}
\newcommand{\intx}{\int\text{d}^3x\ }
\newcommand{\inty}{\int\text{d}^3y\ }
\newcommand{\xp}{\left(\mb{x}'\right)}
\newcommand{\intxp}{\int\text{d}^3x'\ }

\newcommand{\intl}{\int_0^1\text{d}\lambda\ }
\newcommand{\wt}{\widetilde}
\newcommand{\uq}{_\mb{q}}
\newcommand{\uqs}{_{\mb{q}\sigma}}

\newcommand{\uqt}{_{\mb{q}\tau}}
\newcommand{\uqpm}{_{\mb{q}\pm}}
\newcommand{\gee}{\textsl{g}}

\begin{document}
	
	\title{Theory of photon condensation in an arbitrary-gauge condensed matter cavity model}
	
	\author{Dominic M. Rouse}
	\email[]{dominic.rouse@manchester.ac.uk}
	\affiliation{Department of Physics and Astronomy, University of Manchester, Oxford Road, Manchester M13 9PL, United Kingdom}
	
	\author{Adam Stokes}
	\email[]{adamstokes8@gmail.com}
	\affiliation{School of Mathematics, Statistics and Physics, Newcastle University, Newcastle upon Tyne NE1 7RU, United Kingdom}
	\affiliation{Department of Physics and Astronomy, University of Manchester, Oxford Road, Manchester M13 9PL, United Kingdom}
	
	\author{Ahsan Nazir}
	\email[]{ahsan.nazir@manchester.ac.uk}
	\affiliation{Department of Physics and Astronomy, University of Manchester, Oxford Road, Manchester M13 9PL, United Kingdom}
	
	\date{\today}
	
	\begin{abstract}
		We derive an arbitrary-gauge criterion under which condensed matter within an electromagnetic field may transition to a photon condensed phase. 
		Previous results are recovered by selecting the Coulomb-gauge wherein photon condensation can only occur for a spatially-varying field and can be interpreted as a magnetic instability. We demonstrate the gauge-invariance of our description directly, but since matter and photons are gauge-relative concepts we find more generally that photon condensation can occur within a spatially uniform field, and that the relative extent to which the instability is both magnetic and electric versus purely magnetic depends on the gauge. 
	\end{abstract}
	
	\maketitle
	
The discussion of photon condensation in the ground state of the simplest quantum model of many dipoles interacting with a single-mode cavity, the so-called Dicke model, possesses a substantial history  \cite{hepp1973superradiant,wang1973phase,hepp1973equilibrium,carmichael1973higher,rzazewski1975phase,keeling2007coulomb,stokes2020uniqueness,vukics_fundamental_2015,bamba_superradiant_2016,vukics_elimination_2014,grieser_depolarization_2016,vukics_adequacy_2012,bialynicki-birula_no-go_1979,lee_first-order_2004,bamba_circuit_2017,rzazewski_stability_1991,knight_are_1978,emeljanov_appearance_1976,emary_quantum_2003,viehmann_superradiant_2011,rzazewski_remark_1976,pimentel_phase_1975,kudenko_interatomic_1975,gegg_superradiant_2018,nataf_no-go_2010,sung_phase_1979,andolina2019cavity,andolina2020theory,stokes2020implications,Bamba2022,roman-roche_effective_2022}. Myriad `no-go' and `counter no-go' theorems respectively forbid and permit a phase transition into a photon condensed state. Reconciliation of these apparently incompatible results is found \cite{stokes2020uniqueness} by noting that ``light" and ``matter" as quantum subsystems, are gauge-relative \cite{stokes2020implications}. A ``photon", defined as a quantum of ``light", refers to different physical degrees of freedom in each different gauge. The phase of the Dicke model originally discovered by Hepp and Leib \cite{hepp1973superradiant}, can be understood as a condensate of photons defined relative to the multipolar-gauge. The Coulomb-gauge defines the same physical excitations as purely material, such that photon condensation does not occur, and this result is nothing but the well-known ``no-go theorem".

Direct demonstration that the extensively studied criticality of the Dicke model is consistent with gauge-invariance is clearly an important theoretical result, but the physical validity of such a simple model for describing realistic experimental situations is much less clear. A number of approximations with (at least partially) opposing conditions of validity must be simultaneously made in order to coerce the fundamental light-matter Hamiltonian into the form of a Dicke model. In particular, the material system is supposed to be comprised of individually discernible (disjoint) dipoles that form a dense gas, but the spatial dependence of each dipole's coupling to the cavity field is ignored. A physical analysis of the limitations this places upon the light-matter coupling strength was performed by Vukics {\em et al.} in Ref.~\cite{vukics_fundamental_2015}. It was found that the critical point of the phase transition is at the border of
covalent molecule formation and crystallization.

Condensed matter theory naturally possesses a wide variety of phenomena associated with strong correlations \cite{PhysRevLett.122.133602,PhysRevResearch.2.013143,PhysRevB.101.205140,PhysRevX.10.041027,PhysRevLett.125.217402,Rokaj_2018,PhysRevA.98.043801,doi:10.1073/pnas.1814178116,doi:10.1021/acsphotonics.9b01649,PhysRevLett.122.167002,PhysRevB.99.020504,PhysRevB.99.085116,PhysRevB.99.099907,Bamba2022,roman-roche_effective_2022,rokaj_free_2022,eckhardt_quantum_2022,manzanares_superradiant_2022,guerci_superradiant_2020,nataf_rashba_2019,schlawin_cavity_2022,roman-roche_photon_2021}, including superconductivity, excitonic condensation, magnetism and magnonic phase transitions. The difficulties regarding the physical appropriateness of the Dicke model for describing actual experiments are circumvented by more recent analyses of photon condensation in condensed matter cavity QED systems. 
For example, rather than considering individually discernible charge distributions such as dipoles, Ref.~\cite{andolina2020theory} considers a so-called jellium of (strongly correlated) electrons that are superimposed on a uniform positive background. The spatial variation of the cavity field over the extent of the jellium system need not be neglected, and truncation of the material Hilbert space can also be avoided. 

Despite these advantages, the description of critical phenomena within condensed matter cavity QED has so far been confined almost exclusively to the Coulomb-gauge. Photon condensation has been found possible for a spatially varying field, but not for a uniform field, and it has been interpreted as a magnetic instability. In view of the fundamental gauge-relativity exemplified by the criticality of the Dicke model, it is clear that a more general, arbitrary-gauge theory is required to determine whether the same understanding of criticality persists when the restriction to the Coulomb-gauge is lifted.

	In this work, we derive an arbitrary-gauge criterion for photon condensation in the jellium model. 
	The results of Refs.~\cite{andolina2019cavity,andolina2020theory} are reproduced as special cases obtained by choosing the Coulomb-gauge. More generally however, we find that photon condensation can occur through a combination of magnetic and electric instabilities, with the relative contributions of each depending on the gauge. We find moreover, that condensation can generally occur for a uniform field. In particular, correct to the first (electric dipole) order of a multipole expansion, photon condensation within the multipolar-gauge constitutes a purely electric instability. This offers the most extreme alternative to the Coulomb-gauge’s purely magnetic classification of photon condensation, which occurs for a spatially varying field. We conclude with a discussion of the physical significance of our findings.\\
	
{\emph{Hamiltonian}}.---The jellium model consists of $N$ electrons of mass $m$ and charge $-e$ distributed over a homogeneous background of total charge $+Ne$ with volume $V=L^3$ centered at the origin. The charge density operator is $\hat{\rho}\x=\hat{\rho}_e\x+\hat{\rho}_b$ with electron and background parts $\hat{\rho}_e\x=-e\sum_{\mu=1}^N\delta(\mb{x}-\hat{\mb{r}}_\mu)$ and $\hat{\rho}_b=eN/V$, where $\hat{\mb{r}}_\mu$ are the electronic positions. The electron density is $\hat{n}_e=\hat{\rho}_e/(-e)$.

	The energy of the jellium contained within an electromagnetic cavity of volume $V$ is
	\begin{equation}\label{eq:Harb}
	  \hat{H}=\sum_\mu \frac{1}{2}m\dot{\hat{\mb{r}}}_\mu^2+U+\frac{1}{2}\int_V d^3x\left[\hat{\mb{E}}_\text{T}\x^2+\hat{\mb{B}}\x^2\right],
	\end{equation}
	where $\mb{B}$ and $\mb{E}_\text{T}$ are the magnetic and transverse electric fields, and the electrostatic potential is $\hat{U}=\intx\hat{\mb{E}}_\text{L}\x^2/2$. 
	We employ a general formulation in which the choice of gauge is encoded into the arbitrary transverse component ${\bf g}_\text{T}$ of the Green's function for the divergence operator, defined by $\bm{\nabla}_\mb{x}\cdot\mb{g}(\mb{x},\mb{x}')=\delta(\mb{x}-\mb{x}')$. The longitudinal component is fixed uniquely as ${\bf g}_{\rm L}({\bf x},{\bf x}')= -\nabla(4\pi|{\bf x}-{\bf x}'|)^{-1}$ in terms of which Gauss' law yields the static field $\hat{{\bf E}}_{\rm L}({\bf x}) := \int d^3 x' {\bf g}_{\rm L}({\bf x},{\bf x}')\hat{\rho}({\bf x}')$. The arbitrary vector potential and transverse polarisation field 
		\begin{subequations}\label{eq:APpartial}
		\begin{align}
		&\hat{\mb{A}}\x=\hat{\mb{A}}_\text{T}\x+\bm{\nabla}\intxp\mb{g}_\text{T}(\mb{x}',\mb{x})\cdot\hat{\mb{A}}_\text{T}\xp,\label{eq:A}\\
		&\hat{\mb{P}}_\text{T}\x=-\intxp\mb{g}_\text{T}(\mb{x},\mb{x}')\hat{\rho}\xp\label{eq:PT},
		\end{align}
	\end{subequations}
determine the theory's canonical momenta in terms of $m\dot{ \hat{{\bf r}}}_\mu$ and $\hat{{\bf E}}_{\rm T}$ as
	\begin{subequations}\label{eq:pPi}
		\begin{align}
		\hat{\mb{p}}_\mu&=m\dot{\hat{\mb{r}}}_\mu-e\hat{\mb{A}}(\hat{\mb{r}}_\mu),\\ \hat{\bm{\Pi}}\x&=-\hat{\mb{E}}_\text{T}\x-\hat{\mb{P}}_\text{T}\x = -\hat{\mb{E}}\x-\hat{\mb{P}}_g\x, \label{eq:Pi}
		\end{align} 
	\end{subequations}
where $\hat{{\bf P}}_g({\bf x}) := -\int d^3 x' {\bf g}({\bf x},{\bf x}')\hat{\rho}({\bf x}')=\hat{{\bf P}}_\text{T}({\bf x})+\hat{{\bf P}}_\text{L}({\bf x})$ is the total polarisation field in which $\hat{{\bf P}}_\text{L}=-\hat{{\bf E}}_\text{L}$. 

The canonical commutation relations $[\hat{r}_{\mu i},\hat{p}_{\nu j}]=i\delta_{\mu\nu}\delta_{ij}$ and  $[\hat{A}_i\x,\hat{\Pi}_j\xp]=\delta^T_{ij}(\mb{x}-\mb{x}')$ suffice to specify the algebraic properties of all observables. An arbitrary state $\ket{\psi_\mu} $ within the canonical Hilbert space ${\cal H}_\mu$ of the $\mu$'th electron can be expanded in either position or canonical momentum eigenkets as $\ket{\psi_\mu} = \int d^3r_\mu \psi_\mu({\bf r}_\mu)\ket{{\bf r}_\mu} =  \int d^3p_\mu {\tilde \psi}_\mu({\bf p}_\mu)\ket{{\bf p}_\mu}$ where $\psi$ is a square-integrable wave-function and $\tilde \psi$ its Fourier transform. The total material Hilbert space ${\cal H}_m$ is the antisymmetric tensor product of each electronic space; ${\cal H}_m = \bigwedge_{\mu=1}^N {\cal H}_\mu$. 

We assume periodic boundary conditions at the edge of the volume $V$, such that a field ${\bf F}\x$ may be expanded in discrete Fourier modes as $\mb{F}\x=\sum\uq \mb{F}\uq\rme^{i\mb{q}\cdot\mb{x}}$ where $\mb{F}\uq=(1/V)\intx\mb{F}\x\rme^{-i\mb{q}\cdot\mb{x}}$. The annihilation operator for a photon with polarisation $\sigma$ and momentum ${\bf q} = 2\pi(n_x,n_y,n_z)/L$ with $n_i\in\mathbb{Z}$ is defined by $\hat{a}_{{\bf q}\sigma} := {\bm \epsilon}_{{\bf q}\sigma}\cdot [\nu_{\bf q} \hat{{\bf A}}_{\rm T\bf q}+i\hat{{\bf \Pi}}_{\bf q})]/\sqrt{2\nu_{\bf q}}$ where $\nu_{\bf q}=|{\bf q}|$ and the $\bm{\epsilon}\uqs$ with $\sigma =1,2$ are two mutually orthogonal polarisation vectors orthogonal to $\mb{q}$. The eigenstates $\ket{n_{{\bf q}\sigma}}$ of $\hat{a}_{{\bf q}\sigma}^\dagger \hat{a}_{{\bf q}\sigma}$ span the Hilbert space ${\cal H}_{{\bf q}\sigma}$ of the mode ${\bf q}\sigma$, and the photonic Hilbert space is defined as ${\cal H}_{\rm ph}=\bigotimes_{{\bf q}\sigma} {\cal H}_{{\bf q}\sigma}$. The total light-matter Hilbert space is ${\cal H}={\cal H}_m \otimes {\cal H}_{\rm ph}$.
	The canonical operators $\{\hat{{\bf p}}_\mu\}$ and $\hat{{\bf \Pi}}$ represent different observables in each different gauge and so material and photonic states within ${\cal H}_m$ and ${\cal H}_{\rm ph}$ respectively are also physically distinct in each different gauge. We therefore describe the light and matter quantum subsystems as {\em gauge-relative} \cite{stokes2020implications}.
	
Gauss' law, $\nabla\cdot \hat{{\bf E}}=-\nabla \cdot \hat{{\bf P}}=\hat{\rho}$, implies gauge redundancy and the degrees of freedom represented by its solution, $\hat{{\bf P}}_g$, are included within the ``material" quantum subsystem. The remaining part of the electric field is purely photonic. In other words, $\hat{{\bf E}} = -\hat{{\bf \Pi}}-\hat{{\bf P}}_g$ is a partition of the electric field into a purely ``photonic" component, ${\bf \Pi}$, and a purely ``material component", ${\bf P}_g$. The two most commonly chosen gauges of non-relativistic QED, namely the Coulomb- and multipolar-gauges, are given respectively by ${\bf g}_\text{T}={\bf 0}$ and ${\bf g}_\text{T}({\bf x},{\bf x}') = -\int_{\bf o}^{\bf x'} d{\bf z} \cdot \delta^\text{T}({\bf x}-{\bf z})$ where ${\bf z}$ is any curve from a chosen origin ${\bf o}$ to ${\bf x}'$ \cite{stokes2019gauge,stokes2020implications,stokes2020uniqueness}. In the Coulomb-gauge $\hat{{\bf P}}_{\rm T}\equiv {\bf 0}$, therefiore $\hat{{\bf P}}_{\rm Coul}({\bf x})=\hat{{\bf P}}_{\rm L}({\bf x})$ and the photonic field is $\hat{{\bf \Pi}}=-\hat{{\bf E}}_{\rm T}=-\hat{{\bf E}}-\hat{{\bf P}}_{\rm Coul}$. This is a non-local field because $\hat{{\bf E}}$ is local by fundamental assumption while ${\bf P}_{\rm Coul}=-{\bf E}_{\rm L}$ is non-local by definition. The multipolar-gauge 
polarisation $\hat{{\bf P}}_{\rm mult}$ is more localised. By noting that the longitudinal Green's function defining the Coulomb field can be written ${\bf g}_{\rm L}({\bf x},{\bf x}') = {\bf g}_{\rm L}({\bf x},{\bf o}) - \int_{\bf o}^{\bf x'} d{\bf z} \cdot \delta^{\rm L}({\bf x}-{\bf z})$, we see that for a neutral source the multipolar choice of ${\bf g}_{\rm T}$ gives  $\hat{{\bf P}}_{\rm mult}({\bf x})= \int d^3 x' \int_{\bf o}^{\bf x'} d{\bf z} \delta({\bf x}-{\bf z})\hat{\rho}({\bf x}')$, in which each charge is connected along ${\bf z}$ to ${\bf o}$ by a $\delta$-function. By choosing ${\bf o}$ inside the  jellium source, $\hat{{\bf P}}_{\rm mult}$ vanishes outside of the source (which is where the field can be measured), and so $\hat{{\bf \Pi}} = -\hat{{\bf E}}$ at all such points. The multipolar-gauge therefore provides the best possible representation of the local {\em total} electric field $\hat{{\bf E}}$ in terms of a transverse field $\hat{{\bf \Pi}}$ that can be used to define unconstrained photons \cite{stokes2020implications}.
	
	Other noteworthy gauges also exist, such as a gauge in which ground state virtual photon population is highly suppressed as compared to the Coulomb- and multipolar-gauges \cite{stokes2020implications,stokes2019gauge}. Since photons are defined in terms of different gauge-invariant {\em observables} in each different gauge, the physical significance of the condensation of photons defined relative to a particular gauge can only be determined by identifying which observables are accessed by a given detection protocol. Therefore, an analysis that is confined to only one particular gauge such as the Coulomb-gauge, is obviously limited.
	We will derive the condensation criterion with as few restrictions on the form of $\mb{g}_{\rm T}$ as is possible whilst retaining a tractable problem. 
	
	We find it useful for interpreting photon condensation to separate out electric and magnetic interactions using Eqs.~(\ref{eq:APpartial}) as $\hat{H}=\hat{H}_m+\hat{H}_\ell+\hat{H}_E+\hat{H}_B$, where $\hat{H}_m=\sum_\mu \hat{\mb{p}}_\mu^2/(2m)+\hat{U}+\intx\hat{\mb{P}}_\text{T}\x^2/2$ is a purely material Hamiltonian, which includes a $\hat{\mb{P}}_\text{T}^2$ self-term, $\hat{H}_\ell=\intx[\hat{\bm{\Pi}}\x^2+\hat{\mb{B}}\x^2]/2=\sum\uqs\nu\uq\left(\hat{a}^\dagger_{\mb{q}\sigma} \hat{a}_{\mb{q}\sigma}+\frac{1}{2}\right)$ is the bare photonic Hamiltonian, and
	\begin{align}
	\hat{H}_E&=\intx\hat{\mb{P}}_\text{T}\x\cdot\hat{\bm{\Pi}}\x,\label{eq:HE}
	\end{align}
	is the electric interaction Hamiltonian. The magnetic interaction $\hat{H}_B$ is comprised of a paramagnetic component $\hat{H}_B^p=[(e/(2m)]\sum_\mu\{\hat{\mb{p}}_\mu\cdot,\hat{\mb{A}}(\hat{\mb{r}}_\mu)\}$ and a diamagnetic component $\hat{H}_B^d=[e^2/(2m)]\sum_\mu\hat{\mb{A}}(\hat{\mb{r}}_\mu)^2$, which we show in Appendix~\ref{app:HB} can be written in terms of corresponding magnetisation fields to give
	\begin{align}
	\hat{H}_B & =  -\int d^3x \left(\hat{{\bf M}}^p\x+{1\over 2} \hat{{\bf M}}^d \x\right)\cdot\hat{\mb{B}}\x,\label{eq:HB}
	\end{align}
	where $\hat{{\bf M}}_g=\hat{{\bf M}}^p+\hat{{\bf M}}^d$ is the total magnetisation, such that $\hat{{\bf J}}=\partial_t \hat{{\bf P}}_g-\nabla \times \hat{{\bf M}}_g$ is the charge current. 
In terms of photonic operators the Hamiltonian reads
	\begin{equation}\label{eq:Hquant}
	\hat{H}=\hat{H}_m+\sum\uqs A_{\bf q} \bm{\epsilon}\uqs\cdot\left[\hat{\bm{f}}_{\mb{q}}^\dagger \hat{a}\uqs+\hat{\bm{f}}_\mb{q} \hat{a}\uqs^\dagger\right]+\hat{H}_\ell+\hat{H}_B^d,
	\end{equation}
	where $A\uq^2=1/(2\nu\uq V)$ and
	\begin{equation}\label{eq:f}
	\hat{\bm{f}}_{\mb{q}}= iV\nu\uq\left[\left(\check{\mb{q}}\times\hat{\mb{M}}^p_{\mb{q}}\right)+\hat{\mb{P}}_{\text{T}\mb{q}}\right],
	\end{equation}
	with $\check{\mb{q}}={\bf q}/|{\bf q}|$. 
	
	If $\hat{H}_m$ is translationally invariant, then the distribution of electrons is uniform in any eigenstate $\ket{\psi_m}$;
	\begin{align}\label{eq:unif}
		\braket{\psi_m|\hat{n}_e\x|\psi_m} = \frac{N}{V}.
	\end{align}
This property will be used in our analysis below and is proved in Appendix~\ref{app:TIHm}. Translational invariance of ${\hat H}_m$ within the gauges we consider is proved in Appendix~\ref{app:origin}. We note that in the multipolar gauge ${\hat H}_m$ is translationally invariant only under complete translations of both the electrons and the origin ${\bf o}$ to which the electrons are connected by continuous lines of polarisation. Unitary translations of the total material source are accommodated by treating ${\bf o}$ quantum mechanically with conjugate momentum ${\bf p}_{\bf o}$, such that $[o_i,p_{{\bf o}j}]=i\delta_{ij}$ [see Appendix \ref{app:origin}]. Since ${\hat  H}_m$ is independent of ${\bf p}_{\bf o}$ the origin remains fixed. In gauges that do not depend on ${\bf o}$, such as the Coulomb-gauge, the Hamiltonian has the from ${\hat H}_m \equiv {\hat H}_m\otimes I_{\bf o}$ where $I_{\bf o}$ is the identity in the origin sector.

		
{\emph{Thermodynamic Limit}}.---The thermodynamic limit (TDL) is defined by $N,V\to\infty$ with $N/V$ finite. We show in Appendix~\ref{app:TDL} that a generic eigenvector $\ket{\Psi}$ of $\hat{H}$ is separable in the TDL in all gauges; $\ket{\Psi}=\ket{\psi_m}\otimes\ket{\phi_\ell}$ \cite{andolina2019cavity,andolina2020theory}. Photon condensation occurs in the TDL if there exists an eigenstate of $\hat{H}$ containing photons that has a smaller energy than the lowest energy zero-photon state \cite{andolina2019cavity,andolina2020theory}. 
	 We therefore seek the eigenstate $\ket{\phi_\ell}$ of the effective photonic Hamiltonian $\hat{H}_\ell^\text{eff}=\braket{\psi_m|\hat{H}|\psi_m}$  given by
	\begin{align}\label{eq:Hfeff}
	\hat{H}_\ell^\text{eff}=&\ H_m+\sum\uqs A\uq\bm{\epsilon}\uqs\cdot\left[\bm{f}_{\mb{q}}^* \hat{a}\uqs+\bm{f}_\mb{q} \hat{a}\uqs^\dagger\right]\nonumber\\
	&\hspace{2.5cm}+\hat{H}_\ell+\braket{\psi_m|\hat{H}_B^d|\psi_m},
	\end{align}
	where we denote the average of any material operator ${\hat O}_m$ in the state $\ket{\psi_m}$ without a caret; $O_m\equiv\braket{\psi_m|\hat{O}_m|\psi_m}$. 
	
	By expressing the vector potential as $\hat{\mb{A}}\x=\sum\uqs A\uq\left[\bm{\epsilon}'_{-\mb{q}\sigma}\x\hat{a}\uqs+\bm{\epsilon}'\uqs\x\hat{a}\uqs^\dagger\right]$, where
	\begin{equation}\label{eq:polvec}
	\bm{\epsilon}'\uqs(\mb{x})=\bm{\epsilon}\uqs\rme^{-i\mb{q}\cdot\mb{x}}+\bm{\nabla}\intxp\left[\bm{\epsilon}\uqs\cdot\mb{g}_{\rm T}(\mb{x}',\mb{x})\right]\rme^{-i\mb{q}\cdot\mb{x}'},
	\end{equation}
	it becomes clear that determining the eigenstates of $\hat{H}$ remains an intractable problem unless the diamagnetic interaction, $\hat{H}_B^d$, does not couple modes with different momenta. We therefore choose to focus on 
	cases satisfying 
	\begin{equation}\label{eq:ee}
	\braket{\psi_m|\sum_\mu \bm{\epsilon}'\uqs(\hat{\mb{r}}_\mu)\cdot\bm{\epsilon}'_{\mb{q}'\sigma'}(\hat{\mb{r}}_\mu)|\psi_m}\approx ND_{\mb{q}\sigma\sigma'}\delta_{\mb{q},-\mb{q}'},
	\end{equation}
	where $D_{\mb{q}\sigma\sigma'}=D_{\mb{q}\sigma'\sigma}$ is a dimensionless gauge-dependent function. This is sufficient to exemplify the gauge-relativity of photon condensation, and avoids the prescription of a specific model. 
	
	Translational invariance of $\hat{H}_m$ implies Eq.~\eqref{eq:unif}, which in the Coulomb-gauge yields  $D_{\mb{q}\sigma\sigma'}=\delta_{\sigma\sigma'}$ \cite{andolina2020theory}. In the multipolar-gauge, modes of different momenta decouple in the long wavelength limit (LWL) whereby the multipolar diamagnetic interaction vanishes, such that $\lim\{\exp[\pm i\mb{q}.\hat{\mb{r}}_\mu]\to1\}D_{\mb{q}\sigma\sigma'}\to1$ \footnote{Setting field phase factors to unity is the definition of the LWL used throughout this paper.}. More generally, our description applies whenever Eq.~(\ref{eq:ee}) is satisfied, which implies 
	\begin{align}\label{eq:HBd2}
	\braket{\psi_m|\hat{H}_B^d|\psi_m}&=\nonumber\\
	&\hspace{-2cm}\sum_\mb{q}\sum_{\sigma\sigma'}\Delta\uq D_{\mb{q}\sigma\sigma'}\left(\hat{a}_{-\mb{q}\sigma}+\hat{a}_{\mb{q}\sigma}^\dagger\right)\left(\hat{a}_{\mb{q}\sigma'}+\hat{a}_{-\mb{q}\sigma'}^\dagger\right),
	\end{align}
	where $\Delta\uq=(e^2NA\uq^2)/(2m)$. 
	Eq.~\eqref{eq:HBd2} can be diagonalised by introducing new bosonic operators
	\begin{equation}
	\label{eq:bog}
	\hat{c}_{\mb{q}\tau}=w_{\mb{q}\tau}\hat{a}_{\mb{q}1}+x_{\mb{q}\tau}\hat{a}_{\mb{q}2}+y_{\mb{q}\tau}\hat{a}^\dagger_{\mb{q}1}+z_{\mb{q}\tau}\hat{a}^\dagger_{\mb{q}2},
	\end{equation}
	with $\tau\in\{+,-\}$ \cite{qin2001general,de2017virtual}. The transformation results in a displaced oscillator Hamiltonian, $\hat{H}_\ell^\text{eff}=H_m+\sum_{\mb{q}\tau}A\uq\left(g_{\mb{q}\tau}^* \hat{c}_{\mb{q}\tau}+g_{\mb{q}\tau} \hat{c}_{\mb{q}\tau}^\dagger\right)+\sum_{\mb{q}\tau}\nu_{\mb{q}\tau} \left(\hat{c}_{\mb{q}\tau}^\dagger \hat{c}_{\mb{q}\tau}+\frac{1}{2}\right)$. Here $g\uqt=\sum_\sigma h_{\mb{q}\sigma\tau}\left(\bm{\epsilon}_{\mb{q}\sigma}\cdot\bm{f}\uq\right)$, with $h_{\mb{q}1\tau}=w\uqt-y\uqt$ and $h_{\mb{q}2\tau}=x\uqt-z\uqt$, and the renormalised frequency is $\nu\uqt=\nu\uq\lambda\uqt$, where
	\begin{align}\label{eq:lambda}
			\lambda\uqpm=\bigg(1+&\frac{2\Delta\uq}{\nu\uq}\Big[D_{\mb{q}11}+D_{\mb{q}22}\nonumber\\
			&\pm\sqrt{\left[D_{\mb{q}11}-D_{\mb{q}22}\right]^2+4D_{\mb{q}12}^2}\ \Big]\bigg)^{\frac{1}{2}}.
	\end{align} 
	The coefficients in Eq.~(\ref{eq:bog}) are given in Appendix~\ref{app:diag}.

We finally define the diagonal Hamiltonian $\hat{H}_\ell^\text{eff}(\beta)= \hat{D}^\dagger(\beta)\hat{H}^\text{eff}_\ell\hat{D}(\beta)$ where
$\hat{D}(\beta)=\exp[\sum\uqt(\beta\uqt\hat{c}^\dagger\uqt-\beta\uqt^* \hat{c}\uqt)]$ with 	
\begin{align}
	&\beta\uqt=\braket{\psi_m|\hat{\beta}\uqt|\psi_m}, \label{eq:betacon}\\
	&\hat{\beta}\uqt=-\frac{A\uq}{\nu\uqt}\hat{g}\uqt,\label{eq:betahat}
\end{align}
which evaluates as 
	\begin{equation}\label{eq:Hfdisp}
	\hat{H}_\ell^\text{eff}(\beta) = H_m+\sum\uqt\nu\uqt\left(\hat{c}\uqt^\dagger \hat{c}\uqt+\frac{1}{2}-\left|\beta\uqt\right|^2\right).
	\end{equation}
	
	Since the eigenstates of $\hat{H}_\ell^\text{eff}(\beta)$ are number states, $\prod\uqt\ket{\bar{n}\uqt}$, we obtain
	 $\ket{\phi_\ell}=\prod\uqt\ket{\beta\uqt}\equiv\ket{\phi_\ell\{\beta\uqt\}}$ where $\ket{\beta\uqt}=\widehat{D}(\beta)\ket{\bar{n}\uqt}$ is a coherent state. 

{\emph{Condensation criterion}}.---Arbitrarily close to the critical point, a non zero $\braket{\phi_\ell|\hat{a}\uqs|\phi_\ell}$ is signalled by a non-zero $\braket{\phi_\ell|\hat{c}\uqt|\phi_\ell}$, and so $\beta\uqt$ in Eq.~(\ref{eq:betacon}) can be used as the order parameter for the transition \cite{andolina2020theory}. Using Eq.~(\ref{eq:Hfdisp}) the average energy in the state $\ket{\Psi\{\beta\uqt\}}=\ket{\psi_m}\otimes\ket{\phi_\ell\{\beta\uqt\}}$ can then be written
	\begin{equation}\label{eq:Ebeta}
	E_{\bar{n}}\{\beta\uqt\}=H_m+\sum\uqt\nu\uqt\left(\bar{n}\uqt+\frac{1}{2}-\left|\beta\uqt\right|^2\right),
	\end{equation}
where ${\bar n}_{{\bf q}\tau} = \braket{\phi_\ell| \hat{c}^\dagger_{{\bf q}\tau}\hat{c}_{{\bf q}\tau}|\phi_\ell}$. Photon condensation occurs if
	\begin{equation}\label{eq:crit}
	\text{Min}_{\psi_m}\left[E_0(\{\beta\uqt\}\neq 0)\right]<\text{Min}_{\psi_m}\left[E_0(\{\beta\uqt\}=0)\right],
	\end{equation}
	where the minimisation is subject to the constraint defined by Eqs.~(\ref{eq:betacon}) and (\ref{eq:betahat}). 
	Using Eq.~(\ref{eq:Ebeta}), inequality (\ref{eq:crit}) becomes
	\begin{align}\label{eq:stifftext}
	\text{Min}_{\psi_m}&\left[H_m\right]<\braket{\psi^0_m|\hat{H}_m|\psi^0_m}+\sum\uqt\nu\uqt\left|\beta\uqt\right|^2.
	\end{align}

	Constrained minimisation problems of this type can be solved using the stiffness theorem \cite{giuliani2005quantum}, which is derived in Appendix~\ref{app:stiffness} for the case that the constraint involves a spatially varying operator. We begin by defining the zero-temperature static linear response function (SLRF) for material operators $O$ and $C$ by 
	\begin{equation}\label{eq:chiAB}
	\tilde{\chi}_{\mb{q}i,-\mb{q}'j}^{OC}=-2V\sum_{n\neq 0}\frac{\braket{\psi_m^{n}|\hat{O}_{\mb{q}i}|\psi_m^{0}}\braket{\psi_m^{0}|\hat{C}_{-\mb{q}'j}|\psi_m^{n}}}{\varepsilon_m^{(n)}-\varepsilon_m^{(0)}},
	\end{equation}
	where $\hat{H}_m\ket{\psi_m^n}=\varepsilon_m^{(n)}\ket{\psi_m^n}$. In Appendix~\ref{app:TIchi}, we prove that the translational invariance of $\hat{H}_m$ means that the SLRF is also translationally invariant, and so $\tilde{\chi}^{ff}_{\mb{q}i,-\mb{q}'j}=\tilde{\chi}^{ff}_{\mb{q}i,-\mb{q}j}\delta_{\mb{q}\mb{q}'}$. Up to second order in $\delta\beta\uqt=\beta\uqt-\beta_{\mb{q}\tau0}$, where $\beta_{\mb{q}\tau0}=\braket{\psi_m^0|\hat{\beta}\uqt|\psi_m^0}$, one obtains through the stiffness theorem that 
	\begin{equation}\label{eq:critI}
	\text{Min}_{\psi_m}\left[H_m\right]=\braket{\psi_m^0|H_m|\psi_m^0}-\frac{1}{2}V\sum_{\mb{q}\tau}F_{-\mb{q}\tau}\delta\beta\uqt,
	\end{equation}
	where $F\uqt$ is determined by the implicit equation
	\begin{equation}\label{eq:Fs}
	\frac{A\uq^2}{\nu\uqt^2}\tilde{\chi}^{ff}_{\text{T}\mb{q}}\sum_{\tau'}\Lambda_{\mb{q}\tau\tau'}F_{\mb{q}\tau'}-\delta\beta\uqt=0,
	\end{equation}
	in which $\Lambda_{\mb{q}\tau\tau'}=\sum_\sigma h_{\mb{q}\sigma\tau}h_{\mb{q}\sigma\tau'}$ and $\tilde{\chi}^{ff}_{\text{T}\mb{q}}=\sum_{ij}\epsilon_{\mb{q}\sigma i}\epsilon_{\mb{q}\sigma j}\tilde{\chi}_{\mb{q}i,-\mb{q}j}^{ff}$ \cite{giuliani2005quantum}. 
	
 To proceed, we focus on choices of $\mb{g}_\text{T}$ yielding a solution to $F\uqt$ of Eq.~\eqref{eq:Fs} in closed form. This occurs if $\Lambda_{\mb{q}\tau\tau'}\propto\delta_{\tau\tau'}$ which, as we show in Appendix~\ref{app:rotation}, requires that $\hat{H}_m$ be invariant to rotations about $\mb{q}$.    
	As we show in Appendix~\ref{app:stiffness}, rotational invariance implies further that
\begin{align}
\Lambda_{\mb{q}\tau\tau'}=\delta_{\tau\tau'}/\lambda\uqt. \label{eq:taudecoup}
\end{align}

In total we have therefore imposed three restrictions on $\hat{H}_m$, namely, wavevector decoupling in the diamagnetic term [Eq.~(\ref{eq:ee})] and translational and rotational invariance. We note in particular that the Coulomb gauge and the LWL of the multipolar gauge satisfy these restrictions. 
	
 Using Eq.~(\ref{eq:taudecoup}) the solution of Eq.~\eqref{eq:Fs} is found to be $F_{-\mb{q}\tau}=(\delta\beta\uqt^*\nu\uqt^2\lambda\uqt)/(A\uq^2\tilde{\chi}_{\text{T}\mb{q}}^{ff})$. Moreover, Eq.~(\ref{eq:unif}) implies that $\beta_{\mb{q}\tau0}=0$. Using these equalities in Eq.~\eqref{eq:critI} and subsequently in Eq.~\eqref{eq:stifftext} yields,
	\begin{equation}\label{eq:crit0}
	-\sum_{\mb{q}\tau}\left(\frac{V^2\nu\uq^2\lambda\uqt^2 }{\tilde{\chi}^{ff}_{\text{T}\mb{q}}}+1\right)\left|\beta\uqt\right|^2<0.
	\end{equation}
	Since we have optimised the parameters $\{\beta\uqt\}$ to lower the energy, only terms within the sum in Eq.~(\ref{eq:crit0}) that independently satisfy the inequality will acquire a finite displacement $\beta\uqt\neq0$. We can therefore analyse the criterion for each term separately~\cite{andolina2020theory}. We substitute Eqs.~\eqref{eq:betahat} and \eqref{eq:f} into the summand on the left-hand-side of Eq.~\eqref{eq:crit0}, such that by using  $\bm{\epsilon}\uqs\cdot\left(\check{\mb{q}}\times\mb{M}^p\uq\right)=\left(\bm{\epsilon}_{\mb{q}\sigma'}\cdot\mb{M}^p\uq\right)$ with $\sigma'\neq\sigma$, $\mb{q}\cdot\bm{\epsilon}\uqs=0$, and $\sum_{ij}\epsilon_{\mb{q}\sigma i}\epsilon_{\mb{q}\sigma' j}\tilde{\chi}_{\mb{q}i,-\mb{q}j}^{OC}=\tilde{\chi}^{OC}_{\text{T}\mb{q}}\delta_{\sigma\sigma'}=0$, we arrive at the dimensionless ${\bf g}_\text{T}$-dependent condensation criterion
	\begin{equation}\label{eq:CritT}
	-\wt\chi^{M^pM^p}_{\text{T}\mb{q}}- \wt\chi^{P_\text{T}P_\text{T}}_{\text{T}\mb{q}}>\lambda\uqt^2,
	\end{equation}
	which is the main result of this work. The left-hand-side is the sum of the SLRFs associated with the paramagnetic and the electric interactions of the gauge ${\bf g}_{\rm T}$, and the right-hand-side is a ${\bf g}_\text{T}$-dependent function given in Eq.~\eqref{eq:lambda}.

	
{\emph{Classification of the instability}}.---
Linear response theory can be used to provide a physical interpretation of the criterion. Consider an arbitrary operator ${\hat O}_i$ with equilibrium average, $\langle \hat{O}_i\x\rangle_\text{eq}$, defined at zero temperature using a Hamiltonian $\hat{H}_\text{eq}$. We denote by $\langle \hat{O}_i\x\rangle_\delta$ the average change in ${\hat O}_i$ due to a perturbation of $\hat{H}_{\rm eq}$ in the form $\intx \hat{\mb{C}}\x\cdot \mb{F}(\mb{x})$ where $\hat{\mb{C}}\x$ is some coupling operator and $\mb{F}(\mb{x})$ is the perturbing field. We show in Appendix~\ref{app:LinearResponse} that the Fourier amplitudes within the expansion $\langle \hat{O}_i\x\rangle_\delta=\sum\uq O_{\mb{q}i}^\delta\rme^{i\mb{q}\cdot\mb{x}}$ are given according to linear response theory \cite{giuliani2005quantum} by $O^{\delta}_{\mb{q}i}=\sum_{ j}\tilde{\chi}^{OC}_{\mb{q}i,-\mb{q}j} F_{\mb{q}j}$.
	
	If we now consider the electric interaction $\hat{H}_E$ as a perturbation of $\hat{H}_{\rm eq}=\hat{H}_m$ via the perturbing field ${\bf \Pi}\x$, then the response of $\langle\hat{\mb{P}}_\text{T}\x\rangle$ is found to be
	\begin{equation}
	P_{\text{T}\mb{q}i}^\delta=\sum_j\tilde{\chi}_{\mb{q}i,-\mb{q}j}^{P_\text{T}P_\text{T}}\Pi_{\mb{q}j},
	\end{equation}
	where the transverse part of $\tilde{\chi}^{P_\text{T}P_\text{T}}_{\mb{q}i,-\mb{q}j}$ is the same response function as appears in inequality~\eqref{eq:CritT}. Condensation due to this term in inequality~\eqref{eq:CritT} is therefore the result of an \textit{electric instability}. Similarly, the response of $\langle\hat{{\bf M}}^p\x\rangle$ to the perturbation $\hat{H}_B^p$ with perturbing field $-{\bf B}\x$ is
	\begin{equation}
	M^{p,\delta}_{\mb{q}i}=-\sum_j\tilde{\chi}_{\mb{q}i,-\mb{q}j}^{M^pM^p}B_{\mb{q}j},
	\end{equation}
	where the transverse part of $\tilde{\chi}^{M^pM^p}_{\mb{q}i,-\mb{q}j}$ is the same response function as appears in inequality~\eqref{eq:CritT}. Condensation due to this term in inequality~(\ref{eq:CritT}) is therefore the result of a (para)\textit{magnetic instability}.
	
{\emph{Examples}}.---In this section we evaluate Eq.~\eqref{eq:CritT} by making specific choices of ${\bf g}_\text{T}$. The results of Ref.~\cite{andolina2020theory} are recovered by choosing the Coulomb-gauge ${\bf g}_{\rm T}={\bf 0}$. In this case there is no electric interaction Hamiltonian and wavevectors within the diamagnetic term automatically decouple under the assumption of uniformly distributed charges in the ground state [Eq.~(\ref{eq:unif})]. Moreover, we show in Appendix~\ref{app:mag} that in the Coulomb-gauge one can define the total magnetisation-magnetisation SLRF including both paramagnetic and diamagnetic contributions as $\tilde{\chi}_{\mb{q}i,-\mb{q}j}^{MM}=\tilde{\chi}_{\mb{q}i,-\mb{q}j}^{M^pM^p}-\delta_{ij}\tilde{\chi}_{\mb{q}}^{M^d}$, where $\tilde{\chi}_{\mb{q}}^{M^d}=-(e^2 N)/ (mV\nu\uq^2)$. One can show further that in the Coulomb-gauge, $\lambda\uqt^2=1-\tilde{\chi}^{M^d}\uq$, such that inequality (\ref{eq:CritT}) becomes
	\begin{equation}\label{eq:critCoulomb}
	\text{Coulomb-gauge: }-\tilde{\chi}^{MM}_{\text{T}\mb{q}}>1,
	\end{equation}
	which is the result derived in Ref.~\cite{andolina2020theory}. Condensation within the Coulomb-gauge constitutes a purely magnetic instability.
	
	We can also recover the Coulomb-gauge `no-go' theorem for the case of a spatially uniform field \cite{andolina2019cavity}.  The Hamiltonian is given by Eq.~\eqref{eq:Hquant} in the LWL, such that $\exp[\pm i\mb{q}\cdot\hat{\mb{r}}_\mu]\to1$. Taking this limit within the paramagnetic transverse SLRF of the Coulomb-gauge and 
using the Thomas-Reiche-Kuhn sum rule
	\begin{equation}
	\sum_{n\neq n'}\frac{\left|\braket{\psi_m^{n}|\hat{\mathcal{P}}_i|\psi_m^{n'}}\right|^2}{\varepsilon_m^{(n)}-\varepsilon_m^{(n')}}=\frac{mN}{2},
	\end{equation}
	which holds for an arbitrary material level $n'$, we obtain $
	\ \tilde{\chi}_{\text{T}\mb{q}}^{M^pM^p}=\tilde{\chi}^{M^d}\uq$. Inequality (\ref{eq:CritT}) therefore becomes 
	\begin{equation}\label{eq:critcouleda}
	\text{Coulomb-gauge long wavelength limit: }~~0>1.
	\end{equation}
		
	The opposite extreme of this result is provided by the so-called ``dipole-gauge" defined as the multipolar-gauge within the LWL, which causes all magnetic interactions to disappear. It follows that $D_{\mb{q}\sigma\sigma'}=0$ and so $\lambda\uqt=1$. Inequality (\ref{eq:CritT}) therefore becomes
	\begin{equation}\label{eq:critdipolegauge}
	\text{Dipole-gauge: }-
	\wt\chi^{P_\text{T}P_\text{T}}_{\text{T}\mb{q}}>1,
	\end{equation}
	showing that condensation can occur within this gauge, and that it constitutes a purely electric instability.  Eq.~(\ref{eq:chiAB}) can be used to calculate ${\tilde \chi}^{P_{\rm T}P_{\rm T}}_{{\rm T}{\bf q}}$ in the dipole gauge directly and yields $-V{\tilde \chi}^{P_{\rm T}P_{\rm T}}_{{\rm T}{\bf q}} = \sum_\sigma \bm{\epsilon}\uqs \cdot {\bm \alpha}(0) \cdot \bm{\epsilon}\uqs$ where 
	\begin{align}\label{eq;polalph}
	\alpha_{ij}(\omega):= \sum_{n \neq p} \left [ {d^{0n}_{ i}d_{ j}^{n0} \over \epsilon_m^{(n0)}-\omega}+{d^{0n}_{ j}d_{ i}^{n0} \over \epsilon_m^{(n0)}+\omega} \right]
	\end{align}
	is nothing but the {\em polarisability tensor} of the material ground state with $\epsilon_m^{(n0)}=\epsilon_m^{(n)}-\epsilon_m^{(0)}$ and $d^{0n}_{ i}=-e\sum_\mu \bra{\psi_m^{(0)}}r_{\mu i} \ket{\psi_m^{(n)}}$. The polarisability is central to the study of two-photon processes, Raleigh and Raman scattering, and dispersive energy shifts \cite{craig_molecular_1998}.
	
	Essentially the same results, namely the `no-go' theorem (\ref{eq:critcouleda}) and inequality (\ref{eq:critdipolegauge}) are also found when considering the Dicke model describing a dense gas of dipoles \cite{stokes2020uniqueness}. In this case the underlying Hamiltonian restricted to the dipole-gauge yields the Dicke Hamiltonian without any further approximations beyond those used to obtain the `no-go' theorem from the same starting Hamiltonian restricted to the Coulomb-gauge. In this context inequality (\ref{eq:critdipolegauge}) is nothing but the well-known `counter no-go' theorem and it corresponds to the original Hepp-Leib instability \cite{hepp1973superradiant}. Despite initial appearances, one can show, as in the case of the Dicke model \cite{stokes2020uniqueness}, that the different Coulomb-gauge (no-go) and dipole-gauge (counter no-go) results (\ref{eq:critcouleda}) and (\ref{eq:critdipolegauge}) do not constitute a breakdown of gauge-invariance. Rather, they exemplify {\em gauge-relativity} and they actually constitute a verification that gauge-invariance does hold. To see this note first that in the ground state the average electric field must be static; $\langle {\hat {\bf E}}\rangle_G=\langle {\hat {\bf E}}_{\rm L}\rangle_G$ and $\langle {\hat {\bf E}}_{\rm T}\rangle_G ={\bf 0}$. In Appendix~\ref{eq:ET} we focus on the LWL and verify by direct calculation that one does indeed obtain $\langle \hat{{\bf E}}_\text{T}\rangle_G = {\bf 0}$ in both the Coulomb and dipole gauges. 
In the Coulomb-gauge ${\hat {\bf E}}_{\rm T}=-{\hat {\bf \Pi}}$ is purely photonic, so $\langle {\hat {\bf E}}_{\rm T}\rangle_G ={\bf 0}$ follows immediately from the impossibility of photon condensation [inequality (\ref{eq:critcouleda})].  In the dipole-gauge, $\hat{\mb{E}}_\text{T}=-\hat{\bm{\Pi}}-\hat{\mb{P}}_\text{T}$ and so if an instability corresponding to inequality (\ref{eq:critdipolegauge}) results in a macroscopic average $\langle\hat{\bf{P}}_\text{T}\rangle$, then {\em the same gauge-invariant prediction} $\langle \hat{{\bf E}}_\text{T}\rangle=\bf 0$ implies that photon condensation occurs.

{\emph{Discussion}}.---
The fields ${\hat {\bf P}}_g$ and ${\hat {\bf \Pi}}$ define the components of ${\hat {\bf E}}$ that respectively begin (at time $t=0$) attached and detached from ``matter", as defined relative to the gauge $g$. Suppose that the system is perturbed via the introduction of a polarisable test distribution, $D$, such as a detector dipole in the vicinity of a point ${\bf x}_D$ 
outside the support of the source density $\rho$. 
The predicted response of $D$ to $s$ will depend on how ``matter" is defined. 
Assuming the distributions are localised and disjoint means assuming that ``matter" is such that the initially attached electric fields, ${\hat {\bf P}}_s$ and ${\hat {\bf P}}_D$, have disjoint supports; $\int d^3 x {\hat {\bf P}}_s({\bf x})\cdot {\hat {\bf P}}_D({\bf x})=0$, as in the multipolar gauge. This gauge is used 
in conventional quantum optics to define a photodetector dipole \cite{glauber_quantum_2007}, which therefore registers photons defined relative to the {\em dipole gauge}. 

For a source-field system in the ground state the detector responds to the electric energy density of the source via a dispersive energy shift attributed to the exchange of {\em photons} \cite{craig_molecular_1998,PhysRevA.47.2539,doi:10.1080/01442350802045206,PhysRevA.28.2671,PhysRevA.50.3929}. 
Assuming $D$ in its ground state with isotropic polarisability, one obtains the shift $\Delta E = -\alpha^D(0)\langle{\hat {\bf E}}({\bf x}_D)^2\rangle_G/2$, which produces an attractive force  \cite{doi:10.1080/01442350802045206,PhysRevA.50.3929}. In this expression the average electric energy density is that of the source-field system in the absence of $D$, such that our results regarding photon condensation are directly relevant. The electric field ${\hat {\bf E}}({\bf x}_D)$ coincides with minus the photonic momentum, $-{\hat {\bf \Pi}}({\bf x}_D)$, in the dipole gauge. Its average energy density is calculated in the ground state of the source-field system and is a function of the source polarisability [Eq.~\ref{eq;polalph}], which also determines the occurrence of photon condensation. The simplest example consists of a pair of two-level distributions with transition energies $\omega_s$, $\omega_D$, transition dipole moments ${\bf d}_s$, ${\bf d}_D$, and with isotropic static polarisabilities obtained by assuming for both $s$ and $D$ that $d^{0n}_id^{n0}_j=\delta_{ij}|{\bf d}^{0n}|^2/3$, such that $\alpha^{s,D}_{ij}(0)=\delta_{ij}\alpha^{s,D}(0)$ with $\alpha^{s,D}(0) = 2|{\bf d}_{s,D}|^2 / (3\omega_{s,D})$. The shift $\Delta E$ is then proportional to $-\alpha^s(0)\alpha^D(0)\omega_s\omega_D/[(\omega_s+\omega_D)x_D^6]$ in the near zone, and $-\alpha^s(0)\alpha^D(0)/x_D^7$ in the far zone where retardation results in the well-known (Casimir-Polder) decay $x_D^{-7}$.  We see therefore that the response of $D$ is entirely {\em electric}, but it is {\em not generally electrostatic}, and it is {\em photonic}. Thus, an understanding of photon condensation and it's physical effects cannot be restricted to a consideration of magnetic properties, nor to a consideration of ground state (static) average fields.
{\emph{Conclusion}}.--We have derived a general analytic criterion for photon condensation, inequality~\eqref{eq:CritT}, in an arbitrary-gauge specified by ${\bf g}_\text{T}$. This reproduces previous results as special cases, including the condensation of photons defined relative to the Coulomb-gauge and its characterisation as a purely magnetic instability, as well as both the Coulomb-gauge `no-go' and the dipole-gauge `counter no-go' theorems of the long wavelength limit. Our result clearly demonstrates both the gauge-relativity of photon condensation and the gauge-invariance of physical predictions. We have shown that in general, photon condensation arises from both electric and magnetic interactions, which directly reflects the physical differences between photons defined relative to different gauges. 
	
We thank Gian Marcello Andolina and Alessandro Principi for helpful discussions.

	\clearpage
	\appendix
	\onecolumngrid
	\section{The magnetic interaction}\label{app:HB}
	In this appendix we prove that the magnetic interactions given in the main text, $\hat{H}_B=\hat{H}_B^p+\hat{H}_B^d$, where
	\begin{subequations}\label{eq:Form1}
	\begin{align}
			\hat{H}_B^p&=-\intx\hat{{\bf M}}^p\x\cdot\hat{\mb{B}}\x,\\
			\hat{H}_B^d&=-\frac{1}{2}\intx\hat{{\bf M}^d}\x\cdot\hat{\mb{B}}\x,\label{eq:HBd}
	\end{align}
	\end{subequations}
	can be written in the more conventional forms
	\begin{subequations}\label{eq:Form2}
	\begin{align}
		\hat{H}_B^p&=\frac{e}{2m}\sum_\mu\left\{\hat{\mb{p}}_\mu\ \cdot\ ,\hat{\mb{A}}(\hat{\mb{r}}_\mu)\right\},\\
		\hat{H}_B^d&=\frac{e^2}{2m}\sum_\mu\hat{\mb{A}}(\hat{\mb{r}}_\mu)^2,\label{eq:HBd1}
	\end{align}
	\end{subequations}
	where we have symmetrise the paramagnetic interaction, and we derive explicit formulae for $\hat{\mb{M}}^p\x$ and $\hat{\mb{M}}^d\x$. We will prove this by deriving Eqs.~\eqref{eq:Form1} from Eqs.~\eqref{eq:Form2} by considering the more general form, $\sum_\mu \hat{\mb{V}}_\mu\cdot\hat{\mb{A}}(\hat{\mb{r}}_\mu)$, where $\hat{\mb{V}}_\mu$ is an arbitrary operator depending on $\mu$. In the paramagnetic interaction, $\hat{\mb{V}}_\mu=(e/m)\hat{\mb{p}}_\mu$, and in the diamagnetic interaction, $\hat{\mb{V}}_\mu=(e^2/m)\hat{\mb{A}}(\hat{\mb{r}}_\mu)$. Note that the missing factor of $1/2$ as compared to Eq.~\eqref{eq:HBd1} in the diamagnetic $\hat{\mb{V}}_\mu$ is accounted for by the $1/2$ in Eq.~\eqref{eq:HBd}. This choice will be explained soon.
	
	We define the current, $\hat{\mb{j}}^V\x$, associated with $\hat{\mb{V}}_\mu$ via
	\begin{equation}\label{eq:jV}
		\sum_\mu\hat{\mb{V}}_\mu\cdot\hat{\mb{A}}(\hat{\mb{r}}_\mu)=-\intx\hat{\mb{j}}^V\x\cdot\hat{\mb{A}}\x.
	\end{equation}
	At this point we can explain why the diamagnetic interaction has a factor of $1/2$ compared to the paramagnetic interaction in Eqs.~\eqref{eq:Form1}. Substituting the relevant definition of $\hat{\mb{V}}_\mu$ into Eq.~\eqref{eq:jV} we find the paramagnetic and diamagnetic currents to be,
	\begin{align}
		\hat{\mb{j}}^p\x&=-\frac{e}{2m}\sum_\mu\{\hat{\mb{p}}_\mu,\delta(\mb{x}-\hat{\mb{r}}_\mu)\},\label{eq:jp}\\
		\hat{\mb{j}}^d\x&=-\frac{e^2}{2m}\sum_\mu\{\hat{\mb{A}}(\hat{\mb{r}}_\mu),\delta(\mb{x}-\hat{\mb{r}}_\mu)\},\label{eq:jd}
	\end{align}
	where we have symmetrised the expression by using an anti-commutator. (This is only strictly necessary for the paramagnetic current since $[\hat{\mb{p}}_\mu,\delta(\mb{x}-\hat{\mb{r}}_\mu)]\neq0$.) The sum of these currents gives the total, gauge-invariant current,
	\begin{equation}
	\hat{\mb{j}}\x=\hat{\mb{j}}^p\x+\hat{\mb{j}}^d\x=-\frac{e}{2}\sum_\mu\left\{\frac{\hat{\mb{p}}_\mu+e\hat{\mb{A}}(\hat{\mb{r}}_\mu)}{m},\delta(\mb{x}-\hat{\mb{r}}_\mu)\right\}=-\frac{e}{2}\sum_\mu\{\dot{\hat{\mb{r}}},\delta(\mb{x}-\hat{\mb{r}}_\mu)\}.
	\end{equation}
	Had we included the $1/2$ within the definition of $\hat{\mb{V}}_\mu$, and so within $\hat{\mb{j}}^d\x$, the sum of the para- and dia-magnetic currents would not equal the physical current.
	
	We now proceed with the proof by substituting Eq.~\eqref{eq:A} for the total vector potential into Eq.~\eqref{eq:jV} to obtain
	\begin{equation}\label{eq:step1}
		\sum_\mu\hat{\mb{V}}_\mu\cdot\hat{\mb{A}}(\hat{\mb{r}}_\mu)=-\intx\hat{\mb{j}}^V\x\cdot\hat{\mb{A}}_\text{T}\x-\intx\left[\hat{\mb{j}}^V\x\cdot\bm{ \nabla}_\mb{x}\right]\intxp\mb{g}_\text{T}(\mb{x}',\mb{x})\cdot\hat{\mb{A}}_\text{T}({\bf x}').
	\end{equation}
	To move this expression into the form seen in Eqs.~\eqref{eq:Form1} we will use three formulae. The first is found by inverting $\hat{\mb{A}}_\text{T}\x=\bm{\nabla}\times\hat{\mb{B}}\x$ via Helmholtz' theorem as
	\begin{equation}\label{eq:Id1}
		\hat{\mb{A}}_\text{T}\x=\inty\frac{\bm{\nabla}_{\mb{y}}\times\hat{\mb{B}}\y}{4\pi\left|\mb{x}-\mb{y}\right|}.
	\end{equation}
	The second, is the following identity valid for any suitably well-behaved operator field $\hat{\mb{W}}\y$ that vanishes at the boundary of integration,
	\begin{equation}\label{eq:Id2}
		\inty\frac{\bm{\nabla}_{\mb{y}}\times\hat{\mb{W}}\y}{4\pi\left|\mb{x}-\mb{y}\right|}=\inty\bm{\nabla}_{\mb{x}}\times\left[\frac{\hat{\mb{W}}\y}{4\pi\left|\mb{x}-\mb{y}\right|}\right].
	\end{equation}
	Similarly, the third is,
	\begin{equation}\label{eq:Id3}
		\inty\hat{\mb{W}}\y\cdot\left[\bm{\nabla}\times\hat{\mb{Z}}\y\right]=\inty\hat{\mb{Z}}\y\cdot\left[\bm{\nabla}\times\hat{\mb{W}}\y\right].
	\end{equation}
	Using Eqs.~\eqref{eq:Id1}, \eqref{eq:Id2} and \eqref{eq:Id3}, in that order, we can rewrite the first term of Eq.~\eqref{eq:step1} as
	\begin{equation}
		-\intx\hat{\mb{j}}^V\x\cdot\hat{\mb{A}}_\text{T}\x=-\inty\left(\intx\frac{\bm{\nabla}_\mb{x}\times\hat{\mb{j}}^V\x}{4\pi\left|\mb{x}-\mb{y}\right|}\right)\cdot\hat{\mb{B}}\y\equiv-\inty \hat{\mb{M}}^V_0\y\cdot\hat{\mb{B}}\y,
	\end{equation}
	where we have defined the $\mb{g}_\text{T} \equiv {\bf 0}$ component of the magnetisation associated with current $\hat{\mb{j}}^V\y$ as,
	\begin{equation}\label{eq:MV0}
		\hat{\mb{M}}^V_0\y=\intx\frac{\bm{\nabla}_\mb{x}\times\hat{\mb{j}}^V\x}{4\pi\left|\mb{x}-\mb{y}\right|}.
	\end{equation}
	We now move onto the second term in Eq.~\eqref{eq:step1}. After again using Eqs.~\eqref{eq:Id1}, \eqref{eq:Id2} and \eqref{eq:Id3} we find that
	\begin{equation}
		-\intx\left[\hat{\mb{j}}^V\x\cdot\bm{ \nabla}_\mb{x}\right]\intxp\mb{g}_\text{T}(\mb{x}',\mb{x})\cdot\hat{\mb{A}}_\text{T}({\bf x}')\equiv-\inty \hat{\mb{M}}_{\gee_\text{T}}^V\y\cdot\hat{\mb{B}}\y,
	\end{equation}
 where the $\mb{g}_\text{T}$-dependent part of the magnetisation associated with $\hat{\mb{j}}^V\y$ is, 
 \begin{equation}
 	\hat{\mb{M}}_{\gee_\text{T}}^V\y=\intx\left[\hat{\mb{j}}^V\x\cdot\bm{\nabla}_\mb{x}\right]\intxp\frac{\bm{\nabla}_{\mb{x}'}\times\mb{g}_\text{T}(\mb{x}',\mb{x})}{4\pi\left|\mb{x}'-\mb{y}\right|}.
 \end{equation}
This can be brought into a more useful form using integration by parts once more: $\intx \left[\hat{\mb{W}}\x\cdot\bm{\nabla}\right]\mb{Z}\x=-\intx\mb{Z}\x\left[\bm{\nabla}\cdot\hat{\mb{W}}\x\right]$, and subsequently using the relation $\partial_t\hat{\rho}^V\x=-\bm{\nabla}\cdot\hat{\mb{j}}^V\x$. This gives
 \begin{equation}\label{eq:MVg}
	\hat{\mb{M}}_{\gee_\text{T}}^V\y=-\int d^3x'\frac{\bm{\nabla}_{\mb{x}'}\times\partial_t{\hat{\mb{P}}}_\text{T}^V(\mb{x}')}{4\pi\left|\mb{x}'-\mb{y}\right|},
\end{equation}
where we have defined the transverse polarisation associated with charge density $\hat{\rho}^V\x$ as
\begin{equation}\label{eq:PTV}
	\hat{\mb{P}}_\text{T}^V\x=-\intxp \mb{g}_\text{T}(\mb{x},\mb{x}')\hat{\rho}^V\xp.
\end{equation}

Collecting terms, we have therefore shown that
\begin{equation}
	\sum_\mu\hat{\mb{V}}_\mu\cdot\hat{\mb{A}}(\hat{\mb{r}}_\mu)=-\intx \hat{\mb{M}}^V\x\cdot\hat{\mb{B}}\x,
\end{equation}
where $\hat{\mb{M}}^V\x=\hat{\mb{M}}^V_0\x+\hat{\mb{M}}^V_{\gee_\text{T}}\x$ with the contributions given in Eqs.~\eqref{eq:MV0} and \eqref{eq:MVg}. The paramagnetic interaction is \begin{equation}
	\frac{e}{m}\sum_\mu\hat{\mb{p}}_\mu\cdot\hat{\mb{A}}(\hat{\mb{r}}_\mu)=-\intx \hat{\mb{M}}^p\x\cdot\hat{\mb{B}}\x,
\end{equation}
where $\hat{\mb{M}}^p\x$ is defined completely by $\hat{\mb{j}}^p\x$ in Eq.~\eqref{eq:jp}. The diamagnetic interaction is
\begin{equation}\label{eq:HBdia}
	\frac{e^2}{2m}\sum_\mu\hat{\mb{A}}(\hat{\mb{r}}_\mu)\cdot\hat{\mb{A}}(\hat{\mb{r}}_\mu)=-\frac{1}{2}\intx \hat{\mb{M}}^d\x\cdot\hat{\mb{B}}\x,
\end{equation}
where $\hat{\mb{M}}^d\x$ is defined completely by $\hat{\mb{j}}^d\x$ in Eq.~\eqref{eq:jd}. Note the factor of $1/2$ on the right-hand-side of Eq.~\eqref{eq:HBdia}, which we introduced to ensure that $\hat{\mb{M}}^d\x$ is defined by the correct diamagnetic current. 

\section{Proof of Equation~\eqref{eq:unif}}\label{app:TIHm}
	
In this Appendix we prove that Eq.~\eqref{eq:unif} follows from the translational invariance of $\hat{H}_m$. Repeated here, Eq.~\eqref{eq:unif} is
\begin{align}
	\braket{\psi_m|\hat{n}_e\x|\psi_m} = \frac{N}{V},
\end{align}
for any matter eigenstate $\ket{\psi_m}$ of $\hat{H}_m$.
	
We begin by writing the discrete component of $\hat{H}_m$ as
\begin{align}
	\hat{H}_m = \sum_n \sum_{i=1}^{d(n)} E_n\ket{E_n^{(i)}}\bra{E_n^{(i)}},
\end{align}
where the $E_n$ are distinct eigenvalues for different $n$ and is assumed to possess a (possibly infinite) $d(n)$-fold degeneracy. The eigenstates are orthonormal and complete, such that identity can be resolved as
\begin{equation}
	\hat{I}=  \sum_n \sum_{i=1}^{d(n)} \ket{E_n^{(i)}}\bra{E_n^{(i)}},
\end{equation}
with $\braket{E_n^{(i)}|E_m^{(j)}}=\delta_{nm}\delta_{ij}$.  Since $\hat{H}_m$ is translationally invariant, $\hat{T}({\bf a})\hat{H}_m\hat{T}({\bf a})^\dagger =\hat{H}_m$ where $\hat{T}(a)$ is a unitary operator that translates the jellium system by a vector $\mb{a}$. Since $\hat{T}({\bf a})\hat{H}_m = \hat{H}_m\hat{T}({\bf a})$ we have
\begin{align}
	0= \bra{E_n^{(i)}}[\hat{H}_m,\hat{T}({\bf a})]\ket{E_m^{(j)}} = (E_n-E_m)\bra{E_n^{(i)}}\hat{T}({\bf a})\ket{E_m^{(j)}}.
\end{align}
Therefore,
\begin{align}
	t_{nm}^{ij}({\bf a}):=\bra{E_n^{(i)}}\hat{T}({\bf a})\ket{E_m^{(j)}} = \delta_{nm}t^{ij}_{nn}({\bf a}),
\end{align}
and so
\begin{align}
	\hat{T}({\bf a}) =\sum_n \sum_{i,j=1}^{d(n)} t_{nn}^{ij}({\bf a})  \ket{E_n^{(i)}}\bra{E_n^{(j)}}.
\end{align}
It follows that $\hat{T}({\bf a})$ cannot couple different eigenspaces of $H_m$.

We can diagonalise $\hat{T}({\bf a})$ by defining a basis
\begin{align}
	\ket{\epsilon_n^{(p)}} = \sum_{i=1}^{d(n)} S_n^{ip}\ket{E_n^{(i)}},\qquad p=1,...,d(n),
\end{align}
where $\hat{S}_n\hat{S}_n^\dagger = \hat{I}_n = \hat{S}_n^\dagger \hat{S}_n$ for all $n$ with $\hat{I}_n$ the identity within the $n$'th eigenspace. Unitarity of $\hat{S}_n$ implies that the new states are an orthonormal basis and so we may resolve the identity as
\begin{equation}
	\hat{I}=  \sum_n \sum_{p=1}^{d(n)} \ket{\epsilon_n^{(p)}}\bra{\epsilon_n^{(p)}},
\end{equation}
with 
$ \braket{\epsilon_n^{(p)}|\epsilon_m^{(q)}}=\delta_{nm}\delta_{qp}$. We choose $\hat{S}_n$ to diagonalise the matrix $[t_{nn}^{ij}({\bf a})]$ so that
\begin{align}
	\hat{T}({\bf a}) = \sum_n \sum_{p=1}^{d(n)}  t_{nn}^{pp}({\bf a})\ket{\epsilon_n^{(p)}}\bra{\epsilon_n^{(p)}}.
\end{align}
Finally, since $\hat{T}(-{\bf x})^\dagger = \hat{T}({\bf x})$, $\hat{T}({\bf x})\hat{\rho}_e({\bf x})T({\bf x})^\dagger = \hat{\rho}_e({\bf 0})$, and $|t_{nn}^{pp}(\mb{a})|^2=1$, it follows that
\begin{align}
	-eN =& \int d^3 x \bra{\epsilon_n^{(p)}}\hat{\rho}_e({\bf x})\ket{\epsilon_n^{(p)}} = \int d^3 x  \bra{\epsilon_n^{(p)}}\hat{T}(-{\bf x})^\dagger \hat{\rho}_e({\bf x})\hat{T}(-{\bf x})\ket{\epsilon_n^{(p)}} =\int d^3 x  \bra{\epsilon_n^{(p)}}\hat{\rho}_e({\bf 0})\ket{\epsilon_n^{(p)}} \nonumber \\  =& V \bra{\epsilon_n^{(p)}} \hat{\rho}_e({\bf 0})\ket{\epsilon_n^{(p)}} = V\bra{\epsilon_n^{(p)}}\hat{T}(-{\bf x})^\dagger \hat{\rho}_e({\bf x})\hat{T}(-{\bf x})\ket{\epsilon_n^{(p)}} = V \bra{\epsilon_n^{(p)}}\hat{\rho}_e({\bf x})\ket{\epsilon_n^{(p)}},
\end{align}
for any $n$ and $p$. This completes the proof.

		\section{The extended matter Hilbert space}\label{app:origin}
	
	In this appendix we show that $\hat{H}_m$ is translationally invariant in the multipolar-gauge. We note that $\hat{H}_m$ is trivially translationally invariant in the Coulomb gauge since $\mb{P}_\text{T}=\mb{0}$. In the multipolar-gauge, we here explain how extending the matter Hilbert space to include a wave-mechanical quantised origin permits the correct, translationally invariant description of $\hat{H}_m$ in the multipolar-gauge. That $\hat{H}_m$ should be translationally invariant in all gauges is evident; moving the material source by the same vector should not change the physics of the system. Recall from the main text that 
	\begin{equation}\label{eq:Hm}
		\hat{H}_m=\sum_\mu \frac{\hat{\mb{p}}_\mu^2}{2m}+\hat{U}+\frac{1}{2}\intx\hat{\mb{P}}_\text{T}\x^2.
		\end{equation}
	
	Consider the polarisation of the $\mu$'th electron referred to an origin ${\bf o}$,
	\begin{align}
		\hat{{\bf P}}_{\rm T}({\bf x},{\bf o},\hat{{\bf r}}_\mu) = -e\int_{\bf o}^{\hat{{\bf r}}_\mu} d{\bf z}\cdot \delta^{\rm T}({\bf x}-{\bf z}) .
	\end{align}
	In the typical multipolar-gauge, the integration path is chosen as the straight line ${\bf z}={\bf o} + \lambda ({\bf r}_\mu-{\bf o})$ with $\lambda \in [0,1]$, to give
	\begin{align}
		\hat{{\bf P}}_{\rm T}({\bf x},{\bf o},\hat{{\bf r}}_\mu) = -e\int_0^1 d\lambda (\hat{{\bf r}}_\mu-{\bf o}) \cdot \delta^{\rm T}\big({\bf x}-{\bf o}-\lambda (\hat{{\bf r}}_\mu-{\bf o})\big) 
	\end{align}
	From inspection of Eq.~\eqref{eq:PT} this means that
	\begin{align}\label{go}
		{\bf g}_{\rm T}({\bf x},{\bf x}') \equiv  {\bf g}_{\rm T}({\bf x},{\bf x}',{\bf o}) \equiv  {\bf g}_{\rm T}({\bf x}-{\bf o},{\bf x}'-{\bf o})= -\int_0^1 d\lambda ({\bf x}'-{\bf o})\cdot \delta^{\rm T}\big({\bf x}-{\bf o}-\lambda({\bf x}'-{\bf o})\big),
	\end{align}
	is the transverse Green's function for the multipolar-gauge, and
	\begin{align}
		{\bf P}_{\rm Tb}({\bf x}) = -{eN\over V}\int d^3 x' {\bf g}_{\rm T}({\bf x},{\bf x}',{\bf o})
	\end{align}
	is the background polarisation.
	
	The electronic translation operator is defined as
	\begin{equation}
		\hat{T}(\bm{a})=\exp[i\hat{\bm{\mathcal{P}}}\cdot\mb{a}],
	\end{equation}
where $\hat{\bm{\mathcal{P}}}=\sum_\mu\hat{\mb{p}}_\mu$ is the total momentum of the electrons. The transverse polarisation is not translated under translations of the electrons alone: $\hat{T}({\bf a})\hat{{\bf P}}_{\rm T}({\bf x})\hat{T}({\bf a})^\dagger \neq \hat{{\bf P}}_{\rm T}({\bf x}-{\bf a})$ where $\hat{{\bf P}}_{\rm T}({\bf x}) = \sum_\mu \hat{{\bf P}}_{\rm T}({\bf x},{\bf o},\hat{{\bf r}}_\mu) +{\bf P}_{\rm Tb}({\bf x})$ is the full polarisation. Physically, the reason for this is clear. The multipolar gauge connects every electron to the origin ${\bf o}$ by lines of polarisation, which will become stretched/compressed unless the origin is translated with the electrons. In other words, the electronic translation operator alone does not implement translations of the total material source in the multipolar gauge. However, since $\hat{{\bf P}}_{\rm T}({\bf x},{\bf o},\hat{{\bf r}}_\mu)$ depends only on the differences $\hat{{\bf r}}_\mu-{\bf o}$ and ${\bf x}-{\bf o}$, i.e., $\hat{{\bf P}}_{\rm T}({\bf x},{\bf o},\hat{{\bf r}}_\mu)\equiv \hat{{\bf P}}_{\rm T}({\bf x}-{\bf o},{\bf 0},\hat{{\bf r}}_\mu-{\bf o}) $, if one translates both $\hat{{\bf r}}_\mu$ and ${\bf o}$ by ${\bf a}$ then the electronic polarisation will be translated by ${\bf a}$. Therefore $\hat{H}_m$ is invariant with respect to translations of the total material system that includes the origin and associated lines of polarisation.
	
	In order to implement complete translations via a unitary operator, we define an extended jellium model with the Hilbert space ${\cal H}_m\otimes {\cal H}_{\bf o}$ which now includes the Hilbert space ${\cal H}_{\bf o}$ of the origin $\hat{{\bf o}}$ treated as a wave-mechanical position. The momentum conjugate is $\hat{{\bf p}}_{\bf o}$, and $[\hat{o}_i,\hat{p}_{{\bf o}j}]=i\delta_{ij}$. The Hamiltonian $\hat{H}_m =\hat{H}_m(\hat{{\bf o}})$ is given by Eq.~\eqref{eq:Hm}, but is now understood as an operator on the extended Hilbert space. The Hamiltonian $\hat{H}_m$ depends on $\hat{{\bf o}}$ through the polarisation $\hat{{\bf P}}_{\rm T}$, but is independent of $\hat{{\bf p}}_{\rm o}$ and so $\hat{{\bf o}}$ remains fixed and has no affect.
	
	In the extended space, the translation operator is 
	\begin{equation}\label{eq:trans2}
		\hat{T}(\bm{a})=\exp[i\hat{\mathfrak{P}}\cdot\mb{a}],
	\end{equation} 
where $\hat{\mathfrak{P}}=\hat{\mb{p}}_o+\sum_\mu\hat{\mb{p}}_\mu$ now includes the momentum of the origin, $\hat{\mb{p}}_o$, and is such that
	\begin{align}
		&\hat{T}({\bf a})\hat{{\bf r}}_\mu \hat{T}({\bf a})^\dagger = \hat{{\bf r}}_\mu +{\bf a},\\
		&\hat{T}({\bf a})\hat{{\bf o}} \hat{T}({\bf a})^\dagger = \hat{{\bf o}} +{\bf a}.
	\end{align}
	This means that
	\begin{align}
		\hat{T}({\bf a})\hat{{\bf P}}_{\rm T}({\bf x},\hat{{\bf o}},\hat{{\bf r}}_\mu)\hat{T}({\bf a})^\dagger \equiv \hat{T}({\bf a})\hat{{\bf P}}_{\rm T}({\bf x}-\hat{{\bf o}},{\bf 0},\hat{{\bf r}}_\mu-\hat{{\bf o}})\hat{T}({\bf a})^\dagger= \hat{{\bf P}}_{\rm T}({\bf x}-\hat{{\bf o}}-{\bf a},{\bf 0},\hat{{\bf r}}_\mu-\hat{{\bf o}})\equiv \hat{{\bf P}}_{\rm T}({\bf x}-{\bf a},\hat{{\bf o}},\hat{{\bf r}}_\mu),
	\end{align}
	and for the background we have
	\begin{align}
		 \hat{T}({\bf a})\hat{{\bf P}}_{\rm Tb}({\bf x})\hat{T}({\bf a})^\dagger &= {eN\over V}\int d^3 x' \int_0^1 d\lambda ({\bf x}'-\hat{{\bf o}}-{\bf a})\cdot \delta^{\rm T}({\bf x}-\hat{{\bf o}}-{\bf a}-\lambda({\bf x}'-\hat{{\bf o}}-{\bf a})) \nonumber \\ &= {eN\over V}\int d^3 x' \int_0^1 d\lambda ({\bf x}'-\hat{{\bf o}})\cdot \delta^{\rm T}({\bf x}-\hat{{\bf o}}-{\bf a}-\lambda({\bf x}'-\hat{{\bf o}})) = {\bf P}_{\rm Tb}({\bf x}-{\bf a}).
	\end{align}
	Hence $\hat{T}({\bf a})\hat{{\bf P}}_{\rm T}({\bf x})\hat{T}({\bf a})^\dagger =\hat{{\bf P}}_{\rm T}({\bf x}-{\bf a})$ and $\hat{T}({\bf a})\hat{H}_m\hat{T}({\bf a})^\dagger =\hat{H}_m$. Therefore, $\hat{H}_m$ is translationally invariant.

	\section{Separability of the total composite state vector in the thermodynamic limit\label{app:TDL}}
	In this Appendix, we prove that in the thermodynamic limit ($N,V\to \infty$ with $N/V$ finite) the composite state vector can be separated into products of matter and light vectors as $\ket{\Psi}=\ket{\psi_m}\otimes\ket{\phi_\ell}$ in all gauges. This was proven in Ref.~\cite{andolina2020theory} for the Coulomb-gauge, and the derivation presented here closely follows that. The quantised Hamiltonian for our arbitrary-gauge model is given in Eq.~\eqref{eq:Hquant}, repeated here for ease of reading, 
	\begin{equation}
	\hat{H}=\hat{H}_m+\hat{H}_{m-\ell}+\hat{H}_\ell,
	\end{equation}
	where 
	\begin{align}
	\hat{H}_m&=\sum_\mu \frac{\hat{\mb{p}}_\mu^2}{2m}+\hat{U}+\frac{1}{2}\intx\hat{\mb{P}}_\text{T}\x^2,\\
	\hat{H}_{m-\ell}&=\sum\uqs A_{\bf q}\bm{\epsilon}\uqs\cdot\left[\hat{\bm{f}}_{\mb{q}}^\dagger \hat{a}\uqs+\hat{\bm{f}}_\mb{q} \hat{a}\uqs^\dagger\right]+\frac{e^2}{2m}\sum_\mu\hat{\mb{A}}(\hat{\mb{r}}_\mu)^2,\\
	\hat{H}_\ell&=\sum\uqs\nu_{\bf q}\left(\hat{a}^\dagger_{\mb{q}\sigma} \hat{a}_{\mb{q}\sigma}+\frac{1}{2}\right),
	\end{align}
	and $\hat{\bm{f}}_{\mb{q}}=  iV\nu\uq[(\check{\mb{q}}\times\hat{\mb{M}}^p\uq)+\hat{\mb{P}}_{\text{T}\mb{q}}]$. As discussed in Ref.~\cite{andolina2020theory} each Hamiltonian contribution must scale linearly with $N$ to give a non-vanishing and non-diverging energy in the thermodynamic limit. First, it is useful to note that $\hat{M}^p\uq\sim\hat{P}_{\text{T}\mb{q}}\sim 1/\nu\uq$ and so $\hat{f}_{\mb{q}}\sim V\sim  N$, and that all summations over the number of charges scales linearly with $N$. It is not as clear how the light operators $\hat{a}\uqs$ and the number of non-negligible light modes, denoted $N_\text{modes}$, scale with $N$. Let us assume that $N_\text{modes}\sim N^s$ for some real constant $s\le 1$, where the upper limit arises so that the vacuum contribution does not dominate the energy in the thermodynamic limit. In order for $\hat{H}_\ell\sim N$ the light operators must scale as $\hat{a}\uqs\sim N^{(1-s)/2}$. This scaling means that $\hat{\mb{A}}(\hat{\mb{r}}_\mu)\sim N^{s/2}$. Substituting this into the two terms of $\hat{H}_{m-\ell}$ gives,
	\begin{align}
	\sum\uqs A_{\bf q}\bm{\epsilon}\uqs\cdot\left[\hat{\bm{f}}_{\mb{q}}^\dagger \hat{a}\uqs+\hat{\bm{f}}_\mb{q} \hat{a}\uqs^\dagger\right]\sim N^{1+\frac{s}{2}},\label{eq:ParaN}\\
	\frac{e^2}{2m}\sum_\mu\hat{\mb{A}}^2(\mb{r}_\mu)\sim N^{1+s}.\label{eq:DiaMagN}
	\end{align}
	If $s<0$, the combined paramagnetic and paraelectric interaction term in Eq.~\eqref{eq:ParaN} (which ultimately causes photon condensation through lowering the energy of the condensate phase) will vanish in the thermodynamic limit. If $s>0$, in the thermodynamic limit the diamagnetic term in Eq.~\eqref{eq:DiaMagN} will dominate the energy. In both cases, the magnitude of the energy-lowering combined paramagnetic and paraelectric term is dominated by the diamagnetic term as $N$ increases. Therefore, in order for a condensate phase to form at large $N$ we require $s=0$, i.e. $N_\text{modes}\sim N^0$ and $\hat{a}\uqs\sim \sqrt{N}$ \cite{andolina2020theory}.
	
	We can now prove disentanglement in the thermodynamic limit assuming $s=0$. Disentanglement requires that both of the following are true,
	\begin{equation}
	\lim_{N\to\infty}\left[\frac{\hat{H}_m}{N},\frac{\hat{H}_{m-\ell}}{N}\right]\to 0\text{ and } \lim_{N\to\infty}\left[\frac{\hat{H}_\ell}{N},\frac{\hat{H}_{m-\ell}}{N}\right]\to 0.
	\end{equation}
	This is easily proven by noting that the following commutators: $\left[ F(\hat{\mb{r}}_\mu),\hat{\mb{p}}_\nu\right]\propto\delta_{\mu\nu}$ for any function $F(\hat{\mb{r}}_\mu)$ and $[ \hat{a}\uqs,\hat{a}_{\mb{q}'\sigma'}^\dagger]=\delta_{\mb{q}\mb{q}'}\delta_{\sigma\sigma'}$, both remove a factor of $N$ scaling. The former does so by removing a summation over the charges, and the latter through removing two light operators. Therefore, all terms dependent on matter-matter and light-light commutators go to zero as $1/N$, whilst matter-light commutators and any commutator involving the background polarisation will vanish.

	\section{Diagonalisation of the effective light Hamiltonian\label{app:diag}}
	In this Appendix, we give details on the Bogoliubov and subsequent displacement transformations that bring the effective light Hamiltonian in Eq.~\eqref{eq:Hfeff} into the diagonalised form in Eq.~\eqref{eq:Hfdisp}. For ease of reading we repeat Eq.~\eqref{eq:Hfeff} here,
	\begin{equation}
	\hat{H}_\ell^\text{eff}=H_m+\sum\uqs A\uq\bm{\epsilon}\uqs\cdot\left[\bm{f}_{\mb{q}}^* \hat{a}\uqs+\bm{f}_\mb{q} \hat{a}\uqs^\dagger\right]+\hat{H}_\ell+\braket{\psi_m|\hat{H}_B^d|\psi_m}.
	\end{equation}
	Recall that under the assumption that the wavevectors decouple for the ground matter state expectation value of the diamagnetic term [the assumption is given explicitly in Eq.~\eqref{eq:ee}] we find
	\begin{equation}
	\braket{\psi_m|\hat{H}_B^d|\psi_m}\approx\sum_\mb{q}\sum_{\sigma\sigma'}\Delta\uq D_{\mb{q}\sigma\sigma'}\left(\hat{a}_{-\mb{q}\sigma}+\hat{a}_{\mb{q}\sigma}^\dagger\right)\left(\hat{a}_{\mb{q}\sigma'}+\hat{a}_{-\mb{q}\sigma'}^\dagger\right).
	\end{equation}
	The Bogoliubov transformation brings the light-only part of the Hamiltonian into diagonalised form. The light-only part is, $\hat{H}_{\ell-o}=\hat{H}_\ell+\braket{\psi_m|\hat{H}_B^d|\psi_m}$ which can be written in symmetric form using the photon operator commutation relations, to find
	\begin{equation}\label{eq:Hbog}
	\hat{H}_{\ell-o}=\sum\uq\left(\hat{\mathfrak{a}}\uq^{\dagger T}\ \hat{\mathfrak{a}}_{-\mb{q}}^T\right) \begin{pmatrix}
	\zeta\uq & \eta\uq \\
	\eta^*\uq & \zeta^*\uq
	\end{pmatrix}
	\begin{pmatrix}
	\hat{\mathfrak{a}}_{-\mb{q}}\\
	\hat{\mathfrak{a}}_{-\mb{q}}^\dagger
	\end{pmatrix},
	\end{equation}
	where $\hat{\mathfrak{a}}\uq=(\hat{a}_{\mb{q}1},\ \hat{a}_{\mb{q}2})^T$, `$T$' is the transpose operation and we have defined,
	\begin{align}
	\zeta=\frac{1}{2}\begin{pmatrix}
	\nu\uq+\alpha\uq & \gamma\uq \\
	\gamma\uq & \nu\uq+\beta\uq
	\end{pmatrix}\text{ and }
	\eta=\frac{1}{2}\begin{pmatrix}
	\alpha\uq & \gamma\uq \\
	\gamma\uq & \beta\uq
	\end{pmatrix},
	\end{align} 
	where $\alpha\uq=2\Delta\uq D_{\mb{q}11}$, $\beta\uq=2\Delta\uq D_{\mb{q}22}$ and $\gamma\uq=2\Delta\uq D_{\mb{q}12}$. 
	
	The diagonalisation procedure is derived in detail in Section III of Ref.~\cite{qin2001general}, here we illustrate the results for our model. The Hamiltonian is diagonalised by a new set of bosons $\hat{\mathfrak{c}}\uq=(\hat{c}_{\mb{q}+}, \hat{c}_{\mb{q}-})^T$ when
	\begin{equation}\label{eq:HbogDiag}
	\hat{H}_{\ell-o}=\sum\uq\left(\hat{\mathfrak{c}}\uq^{\dagger T}\ \hat{\mathfrak{c}}_{-\mb{q}}^T\right) \begin{pmatrix}
	\Omega\uq & 0 \\
	0 & \Omega\uq
	\end{pmatrix}
	\begin{pmatrix}
	\hat{\mathfrak{c}}_{-\mb{q}}\\
	\hat{\mathfrak{c}}_{-\mb{q}}^\dagger
	\end{pmatrix},
	\end{equation}
	where we have defined the diagonal matrix: $\Omega\uq=(1/2)\text{diag}(\nu_{\mb{q}+},\nu_{\mb{q}-})$. The new and old operators are related through
	\begin{equation}
	\begin{pmatrix}
	\hat{\mathfrak{c}}_{-\mb{q}}\\
	\hat{\mathfrak{c}}_{-\mb{q}}^\dagger
	\end{pmatrix}
	=M\uq\begin{pmatrix}
	\hat{\mathfrak{a}}_{-\mb{q}}\\
	\hat{\mathfrak{a}}_{-\mb{q}}^\dagger
	\end{pmatrix}
	=\begin{pmatrix}
	u\uq & v\uq \\
	v\uq^* & u\uq^*
	\end{pmatrix}\begin{pmatrix}
	\hat{\mathfrak{a}}_{-\mb{q}}\\
	\hat{\mathfrak{a}}_{-\mb{q}}^\dagger
	\end{pmatrix},
	\end{equation}
	where 
	\begin{align}
	u\uq=\begin{pmatrix}
	w_{\bm{q}+} & x_{\bm{q}+}  \\
	w_{\bm{q}-}  & x_{\bm{q}-} 
	\end{pmatrix}\text{ and }
	v\uq=\begin{pmatrix}
	y_{\bm{q}+} & z_{\bm{q}+}  \\
	y_{\bm{q}-}  & z_{\bm{q}-} 
	\end{pmatrix}.
	\end{align} 
	One can verify that this matrix representation is identical to Eq.~\eqref{eq:bog}. Enforcing the canonical commutation relations of the new and old boson operators yields the identification that $M^{-1}\uq=KM\uq^\dagger K$ where $K=\text{diag}(I_2,-I_2)$ and $I_2$ is the two dimensional identity matrix. Comparison of Eqs.~\eqref{eq:Hbog} and \eqref{eq:HbogDiag} gives
	\begin{equation}
	\begin{pmatrix}
	\zeta\uq & \eta\uq \\
	\eta^*\uq & \zeta^*\uq
	\end{pmatrix}=M\uq^\dagger \begin{pmatrix}
	\Omega\uq & 0 \\
	0 & \Omega\uq
	\end{pmatrix}M\uq,
	\end{equation}
	and so
	\begin{equation}
	\left(M\uq^\dagger\right)^{-1}\begin{pmatrix}
	\zeta\uq & -\eta\uq \\
	\eta^*\uq & -\zeta^*\uq
	\end{pmatrix}M\uq^\dagger=\begin{pmatrix}
	\Omega\uq & 0 \\
	0 & -\Omega\uq
	\end{pmatrix},
	\end{equation}
	where we have used that $M\uq K M\uq^\dagger=M^\dagger\uq K M\uq=K$. Therefore, the eigenvalues of 
	\begin{equation}
	\begin{pmatrix}
	\zeta\uq & -\eta\uq \\
	\eta^*\uq & -\zeta^*\uq
	\end{pmatrix}
	\end{equation}
	are $\nu_{\mb{q}+}/2$, $\nu_{\mb{q}-}/2$, $-\nu_{\mb{q}+}/2$ and $-\nu_{\mb{q}-}/2$, and the eigenvectors are the column vectors of $M\uq^\dagger$ which correspond to $\hat{c}_{\mb{q}+}^\dagger$, $\hat{c}_{\mb{q}-}^\dagger$, $\hat{c}_{\mb{q}+}$ and $\hat{c}_{\mb{q}-}$. 
	
	After performing the calculation we find that the eigenenergies are $\nu\uqt=\nu\uq\lambda\uqt$ where
	\begin{equation}
	\lambda\uqpm=\bigg(1+\frac{2\Delta\uq}{\nu\uq}\Big[D_{\mb{q}11}+D_{\mb{q}22}\pm\sqrt{\left[D_{\mb{q}11}-D_{\mb{q}22}\right]^2+4D_{\mb{q}12}^2}\ \Big]\bigg)^{\frac{1}{2}},
	\end{equation}
	and the coupling strengths are
	\begin{equation}
	g\uqt=\sum_\sigma h_{\mb{q}\sigma\tau}\left(\bm{\epsilon}_{\mb{q}\sigma}\cdot\bm{f}\uq\right),
	\end{equation}
	where $h_{\mb{q}1\tau}=w\uqt-y\uqt$ and $h_{\mb{q}2\tau}=x\uqt-z\uqt$. The transformation coefficients are
	\begin{equation}\label{eq:bogcoef}
	y\uqt=\frac{1}{\sqrt{N\uqt}}\Phi\uqt,\hspace{1.55cm} z\uqt=\frac{1}{\sqrt{N\uqt}},
	\end{equation}
	$w\uqt=-y\uqt\Theta\uqt$ and $x\uqt=-z\uqt\Theta\uqt$. We have defined $\Theta\uqt=(1+\lambda\uqt)/(1-\lambda\uqt)$, $\Phi\uqpm=d\uq \pm (1+d\uq^2)^{1/2}$ where \begin{equation}\label{eq:d}
		d\uq=\frac{D_{\mb{q}11}-D_{\mb{q}22}}{2D_{\mb{q}12}},
\end{equation} 
and finally the normalisation constants, $N\uqt=8\lambda\uqt\left(1-\lambda\uqt\right)^{-2}\left(1+d\uq\Phi\uqt\right)$.

	After diagonalisation of the light-only Hamiltonian we arrive at
	\begin{equation}\label{eq:Hfapp}
	\hat{H}_\ell^\text{eff}=H_m+\sum_{\mb{q}\tau}A\uq\left(g_{\mb{q}\tau}^* \hat{c}_{\mb{q}\tau}+g_{\mb{q}\tau} \hat{c}_{\mb{q}\tau}^\dagger\right)+\sum_{\mb{q}\tau}\nu_{\mb{q}\tau} \left(\hat{c}_{\mb{q}\tau}^\dagger \hat{c}_{\mb{q}\tau}+\frac{1}{2}\right).
	\end{equation}
	
	Since Eq.~\eqref{eq:Hfapp} is a displaced harmonic oscillator Hamiltonian it is diagonalised with the displacement transformation:
	$\hat{H}_\ell^\text{eff}(\beta)=\hat{D}^\dagger(\beta)\hat{H}^\text{eff}_\ell\hat{D}(\beta)$ where $\hat{D}(\beta)=\exp[\sum\uqt(\beta\uqt\hat{c}^\dagger\uqt-\beta\uqt^* \hat{c}\uqt)]$. The displacement operator transforms the boson operators by $\hat{D}^\dagger(\beta)\hat{c}\uqt\hat{D}(\beta)=\hat{c}\uqt+\beta\uqt$ and so the displaced Hamiltonian is
	\begin{align}
	\hat{H}_\ell^\text{eff}(\beta)=H_m+\sum\uqt\Big(&\nu\uqt\left[\hat{c}^\dagger\uqt\hat{c}\uqt+\frac{1}{2}+\left|\beta\uqt\right|^2\right]+A\uq\left[g^*\uqt\beta\uqt+g\uqt\beta^*\uqt\right]\nonumber\\
	&\qquad+\hat{c}\uqt^\dagger\left[A\uq g\uqt+\beta\uqt\nu\uqt\right]+\hat{c}\uqt\left[A\uq g\uqt^*+\beta\uqt^*\nu\uqt\right]\Big).
	\end{align}
	This is diagonal if the second line vanishes which requires that
	\begin{equation}\label{eq:betaApp}
	\beta\uqt=-\frac{A\uq}{\nu\uqt}g\uqt,
	\end{equation} 
	and so gives the displacement constraint in the main text: $\braket{\psi_m|\hat{\beta}\uqt|\psi_m}=\beta\uqt$ where $\hat{\beta}\uqt$ is given in Eq.~\eqref{eq:betahat}. Using the choice of $\beta\uqt$ in Eq.~\eqref{eq:betaApp} yields the diagonalised effective light Hamiltonian in Eq.~\eqref{eq:Hfdisp}.

	\section{The Stiffness Theorem for a spatially varying constraint\label{app:stiffness}}
	The stiffness theorem allows one to calculate constrained minimisation problems of the form
	\begin{equation}\label{eq:stiff}
	E(\beta)\equiv\text{Min}_{\psi\to\beta}\left[\braket{\psi|\hat{H}|\psi}\right].
	\end{equation}
	This expression indicates that the vector $\ket{\psi}$ is chosen such that it represents the state which gives the smallest expectation value for $H=\braket{\psi|\hat{H}|\psi}$ whilst fulfilling the constraint: $\braket{\psi|\hat{\beta}\uqt|\psi}=\beta\uqt$. In Section 3.2.9 of Ref.~\cite{giuliani2005quantum}, the solution is derived for $\hat{\beta}\uqt$ being an Hermitian operator that does not vary in space, and is there denoted by $\hat{A}$. It is also assumed that $\hat{A}$ has zero ground state expectation value, $A_0\equiv\braket{\psi_0|\hat{A}|\psi_0}=0$, where $\ket{\psi_0}$ is the lowest energy eigenstate of $\hat{H}$. The main 
	difference between the stiffness theorems derived with the constraining operator $\hat{A}$ compared to $\hat{\beta}\uqt$ is that since the spatially varying field $\hat{\beta}_\tau\x=\sum\uq\hat{\beta}\uqt\rme^{i\mb{q}\cdot\mb{x}}$ is Hermitian, $\hat{\beta}\uqt^\dagger=\hat{\beta}_{-\mb{q}\tau}$ is not Hermitian. A second difference, in general, is that the ground state expectation value of $\hat{\beta}\uqt$ may be nonzero. In the case of interest here, however, this average is zero for uniformly distributed charges in the ground state, $\ket{\psi_0}$, and this follows from the translational invariance of $H_m$. We now derive the stiffness theorem for the non-Hermitian constraining operator $\hat{\beta}\uqt$ with non-zero ground state value. 
	The derivation relies on results from linear response theory presented in Appendix~\ref{app:LinearResponse} and derived in detail in Section 3.2 of Ref.~\cite{giuliani2005quantum}.
	
	The stiffness theorem proof begins with an ansatz that the vector solving the minimisation problem in Eq.~\eqref{eq:stiff} is the ground state, $\ket{\psi_\beta}$, of 
	\begin{equation}\label{eq:HA}
	\hat{H}_\beta=\hat{H}+\sum_\tau\intx F_\tau\x \left(\hat{\beta}_\tau\x-\beta_{\tau 0}\x\right),
	\end{equation}
	where $F_\tau\x$ is an undefined field that we will use to ensure that the ansatz is correct, and $\beta_{\tau 0}\x=\braket{\psi_0|\hat{\beta}_\tau\x|\psi_0}$ where $\ket{\psi_0}$ is the ground state of $\hat{H}$. The proof of the correctness of the ansatz has two steps. First we must prove that $\ket{\psi_\beta}$ meets the constraint $\braket{\psi_\beta|\hat{\beta}\uqt|\psi_\beta}=\beta\uqt$; and second, we must prove that $\ket{\psi_\beta}$ is the lowest energy eigenvector of $\hat{H}_\beta$ that does so.
	
	To prove the first step we define a time dependent Hamiltonian,
	\begin{equation}\label{eq:HAt}
	\hat{H}_\beta(t)=\hat{H}+\Theta(t-t_0)\sum_\tau\intx F_\tau(\mb{x},t) \left(\hat{\beta}_\tau\x-\beta_{\tau 0}\x\right),
	\end{equation}
	where $\Theta(t-t_0)$ is the Heaviside step function. For $t<t_0$ the ground state is $\ket{\psi_0}$ and for $t\ge t_0$ it is denoted $\ket{\psi_\beta(t)}$. From Eq.~\eqref{eq:Oresp}, the response of the expectation value of $\hat{\beta}_\tau\x$ away from equilibrium (defined by $\hat{H}$ at zero temperature) due to the time-dependent perturbation in Eq.~\eqref{eq:HAt} is,
	\begin{align}\label{eq:LinRespStiffness}
	\langle \hat{\beta}_\tau(\mb{x},t)\rangle_\delta&\equiv\braket{\psi_\beta(t)|\hat{\beta}_\tau\x|\psi_\beta(t)}-\braket{\psi_0|\hat{\beta}_\tau\x|\psi_0}\\
	&=\sum_{\tau'}\int_0^\infty\text{d}s\ \intxp \zeta^{\beta\beta}_{\tau\tau'}(\mb{x},\mb{x}',s)F_{\tau'}(\mb{x}',t-s).\nonumber
	\end{align}
	As in Appendix~\ref{app:LinearResponse} the time-dependent linear response function is given by
	\begin{equation}
	\zeta_{\tau\tau'}^{\beta\beta}(\mb{x},\mb{x}',s)=-i\braket{\psi_0|\left[\hat{\beta}_\tau(\mb{x},s),\hat{\beta}_{\tau'}\x'\right]|\psi_0}\Theta(s).
	\end{equation}
	Assuming that the perturbing field is static, $F_\tau(\mb{x},t)\to F_\tau(\mb{x})$, the $s$ integral in Eq.~(\ref{eq:LinRespStiffness}) defines the static linear response function,
	\begin{equation}
	\int_0^\infty\text{d}s\ \zeta^{\beta\beta}_{\tau\tau'}(\mb{x},\mb{x}',s)=\zeta^{\beta\beta}_{\tau\tau'}(\mb{x},\mb{x}',\omega=0)\equiv\chi^{\beta\beta}_{\tau\tau'}(\mb{x},\mb{x}').
	\end{equation}
	Inserting $\hat{\beta}_\tau\x=\sum\uq\hat{\beta}\uqt\rme^{i\mb{q}\cdot\mb{x}}$ into Eq.~\eqref{eq:LinRespStiffness} gives
	\begin{equation}
	\sum\uq\left[\beta\uqt-\beta_{\mb{q}\tau 0}\right]\rme^{i\mb{q}\cdot\mb{x}}=\sum\uq\big[V\sum_{\mb{q}'\tau'}\chi_{\mb{q}\tau,-\mb{q}'\tau'}^{\beta\beta}F_{\mb{q}'\tau'}\big]\rme^{i\mb{q}\cdot\mb{x}},
	\end{equation}
	where
	\begin{align}\label{eq:chiFT}
	\chi^{\beta\beta}_{\mb{q}\tau,-\mb{q}'\tau'}=\frac{1}{V^2}&\intx \rme^{-i\mb{q}\cdot\mb{x}}\intxp\rme^{i\mb{q}'\cdot\mb{x}'}\chi^{\beta\beta}_{\tau\tau'}(\mb{x},\mb{x}').
	\end{align}
	Therefore, the ground state of the Hamiltonian in Eq.~\eqref{eq:HA}, $\ket{\psi_\beta}$, will have the correct expectation value of $\hat{\beta}\uqt$ (and so will satisfy the constraint) if the following condition is met by the arbitrary fields $F_\tau\x$,
	\begin{equation}\label{eq:cond}
	\beta\uqt-\beta_{\mb{q}\tau 0}=V\sum_{\mb{q}'\tau'}\chi_{\mb{q}\tau,-\mb{q}'\tau'}^{\beta\beta}F_{\mb{q}'\tau'}.
	\end{equation}
	
	The proof of the second step follows immediately by noting that the counter claim must be false. If another eigenstate $\ket{\psi_\beta'}$ of $\widehat{H}_\beta$ existed which had the correct expectation value of $\widehat{\beta}\uqt$ but a smaller energy eigenvalue of $\widehat{H}$, then $\braket{\psi_\beta'|\widehat{H}|\psi_\beta'} < \braket{\psi_\beta|\widehat{H}|\psi_\beta}$ and this implies that
	\begin{equation}
	\braket{\psi_\beta'|\widehat{H}_\beta|\psi_\beta'} < \braket{\psi_\beta|\widehat{H}_\beta|\psi_\beta},
	\end{equation}
	which is a contradiction; $\ket{\psi_\beta}$ is the ground state of $\widehat{H}_\beta$ by definition \cite{giuliani2005quantum}. 
	
	We have proven that the ground state of Eq.~\eqref{eq:HA}, $\ket{\psi_\beta}$, is the state that solves the minimisation problem within Eq.~\eqref{eq:stiff} if $F_\tau\x$  meets the condition in Eq.~\eqref{eq:cond}. Therefore, we know that the solution to the constrained minimisation problem in Eq.~\eqref{eq:stiff} is
	\begin{align}
	E(\beta)=\braket{\psi_\beta|\hat{H}|\psi_\beta}&=\braket{\psi_\beta|\hat{H}_\beta|\psi_\beta}-\sum_\tau\intx F_\tau\x \left[\beta_\tau\x-\beta_{\tau 0}\x\right]\nonumber\\
	&=\braket{\psi_\beta|\hat{H}_\beta|\psi_\beta}-V\sum_{\mb{q}\tau}F_{-\mb{q}\tau}\left(\beta_{\mb{q}\tau}-\beta_{\mb{q}\tau 0}\right),\label{eq:EbetaApp}
	\end{align} 
	where in the first line we have used  Eq.~\eqref{eq:HA} and in the second used the definition of Fourier transformation. We now need to solve Eq.~\eqref{eq:EbetaApp}.
	
	To evaluate the term $\braket{\psi_\beta|\hat{H}_\beta|\psi_\beta}$ we start by defining another Hamiltonian,
	\begin{equation}
	\hat{H}_\beta(\lambda)=\hat{H}+\lambda V\sum_{\mb{q}\tau}F_{-\mb{q}\tau}\left(\hat{\beta}_{\mb{q}\tau}-\beta_{\mb{q}\tau 0}\right),
	\end{equation}
	whose ground state we denote by $\ket{\psi_\beta(\lambda)}$ [note that $\ket{\psi_\beta(0)}=\ket{\psi_0}$], which has energy $\epsilon(\lambda)=\braket{\psi_\beta(\lambda)|\hat{H}_\beta(\lambda)|\psi_\beta(\lambda)}$. This Hamiltonian is useful because it allows us to make fruitful use of the Hellman-Feynman identity,
	\begin{align}
	\epsilon(1)&=\epsilon(0)+\intl \braket{\psi_\beta(\lambda)|\frac{\partial\hat{H}_\beta(\lambda)}{\partial\lambda}|\psi_\beta(\lambda)}\nonumber\\
	&=E(0)+V\sum_{\mb{q}\tau}F_{-\mb{q}\tau}\intl \braket{\psi_\beta(\lambda)|\left(\hat{\beta}_{\mb{q}\tau}-\beta_{\mb{q}\tau 0}\right)|\psi_\beta(\lambda)},\label{eq:e1}
	\end{align}
	where we have used that $\epsilon(0)=E(0)$. Since $\hat{H}_\beta(\lambda=1)=\hat{H}_\beta$, we know that $\epsilon(1)=\braket{\psi_\beta|\hat{H}_\beta|\psi_\beta}$ and therefore, using Eq.~\eqref{eq:EbetaApp} we know that the solution to Eq.~\eqref{eq:stiff} will be
	\begin{equation}\label{eq:EB}
	E(\beta)=\epsilon(1)-V\sum_{\mb{q}\tau}F_{-\mb{q}\tau}\left(\beta_{\mb{q}\tau}-\beta_{\mb{q}\tau 0}\right).
	\end{equation}
	
	Therefore, to find $E(\beta)$ it remains only to calculate $\epsilon(1)$ in Eq.~\eqref{eq:e1}. To this end we must evaluate the expectation value of $\braket{\psi_\beta(\lambda)|\hat{\beta}_{\mb{q}\tau}|\psi_\beta(\lambda)}$ within the $\lambda$ integral, and to do this we expand $\ket{\psi_\beta(\lambda)}$ in powers of $\lambda$ up to first order as
	\begin{equation}\label{eq:pertex}
	\ket{\psi_\beta(\lambda)}=\ket{\psi_0}-\lambda\sum_{\mb{q}'\tau'} F_{-\mb{q}'\tau'}\sum_{n\neq 0}\ket{\psi_n}\frac{\braket{\psi_n|\widehat{\beta}_{\mb{q}'\tau'}|\psi_0}}{\varepsilon_n-\varepsilon_0}+\mathcal{O}(\lambda^2),
	\end{equation}
	where $\widehat{H}\ket{\psi_n}=\varepsilon_n\ket{\psi_n}$. Note that following this expansion $E(\beta)$ will be found up to second order. Substituting Eq.~(\ref{eq:pertex}) into the expectation value in Eq.~\eqref{eq:e1} gives
	\begin{align}
	\braket{\psi_\beta(\lambda)|\hat{\beta}_{\mb{q}\tau}|\psi_\beta(\lambda)}&=\beta_{\mb{q}\tau 0}-\lambda V\sum_{\mb{q}'\tau'}\sum_{n\neq 0}\Big(F_{-\mb{q}'\tau'}\frac{\braket{\psi_0|\hat{\beta}_{\mb{q}\tau}|\psi_n}\braket{\psi_n|\hat{\beta}_{\mb{q}'\tau'}|\psi_0}}{\varepsilon_n-\varepsilon_0}\nonumber\\
	&\hspace{4cm}+F_{\mb{q}'\tau'}\sum_{n\neq0}\frac{\braket{\psi_0|\hat{\beta}_{-\mb{q}'\tau'}|\psi_n}\braket{\psi_n|\hat{\beta}_{\mb{q}\tau}|\psi_0}}{\varepsilon_n-\varepsilon_0}\Big)+\mathcal{O}\left(\lambda^2\right)\\
	&=\beta_{\mb{q}\tau 0}+\frac{1}{2}\lambda V\sum_{\mb{q}'\tau'}\left[F_{-\mb{q}'\tau'}\chi_{\mb{q}\tau,\mb{q}'\tau'}^{\beta\beta}+F_{\mb{q}'\tau'}\chi_{-\mb{q}'\tau',\mb{q}\tau}^{\beta\beta}\right]+\mathcal{O}\left(\lambda^2\right),
	\end{align}
	where in the second line we have used the zero temperature Lehmann representation of the static linear response functions,
	\begin{equation}
	\chi_{\pm\mb{q}\tau,\pm\mb{q}'\tau'}^{\beta\beta}=-2\sum_{n\neq 0}\frac{\braket{\psi_0|\hat{\beta}_{\pm\mb{q}\tau}|\psi_n}\braket{\psi_n|\hat{\beta}_{\pm\mb{q}'\tau'}|\psi_0}}{\varepsilon_n-\varepsilon_0}.
	\end{equation} 
	Using the reciprocity of static linear response functions, $\chi^{\beta\beta}_{-\mb{q}'\tau',\mb{q}\tau}=\chi^{\beta\beta}_{\mb{q}\tau,-\mb{q}'\tau'}$, performing the $\lambda$ integral, and substituting the resulting expression for $\epsilon(1)$ into Eq.~\eqref{eq:EB}, we arrive at the solution
	\begin{equation}
	E(\beta)=E(0)+\frac{1}{4} V^2\sum_{\mb{q}\tau}F_{-\mb{q}\tau}\sum_{\mb{q}'\tau'}\left[F_{-\mb{q}'\tau'}\chi^{\beta\beta}_{\mb{q}\tau,\mb{q}'\tau'}+F_{\mb{q}'\tau'}\chi^{\beta\beta}_{\mb{q}\tau,-\mb{q}'\tau'}\right]-V\sum_{\mb{q}\tau}F_{-\mb{q}\tau}\left(\beta_{\mb{q}\tau}-\beta_{\mb{q}\tau 0}\right).
	\end{equation}
	Relabelling the summed index ${\bf q}'$ in the first term in the square brackets as $\mb{q}'\to-\mb{q}'$, and then using Eq.~\eqref{eq:cond} gives
	\begin{equation}\label{eq:stiff2}
	E(\beta)=E(0)-\frac{1}{2}V\sum_{\mb{q}\tau}F_{-\mb{q}\tau}\left(\beta_{\mb{q}\tau}-\beta_{\mb{q}\tau 0}\right)
	\end{equation}
	which is the solution to Eq.~\eqref{eq:stiff}. We now need to solve Eq.~\eqref{eq:cond} for $F_{\mb{q}\tau}$ and substitute this into Eq.~\eqref{eq:stiff2}. Before doing so, we note that $F\uqt$ is clearly linear in $\beta\uqt$ and so the second term in Eq.~\eqref{eq:stiff2} is quadratic in $\beta\uqt$.
	
	A general solution to Eq.~\eqref{eq:cond} for $F_{\mb{q}\tau}$ is not forthcoming due to the summations over $\mb{q}$ and $\tau$. We show however that if the static linear response functions are translationally invariant, then we are able to find a closed-form solution in particular cases. First, recall from Eq.~\eqref{eq:betahat} that $\hat{\beta}\uqt=-(A\uq/\nu\uqt)\hat{g}\uqt$ where $\hat{g}\uqt=\sum_\sigma h_{\mb{q}\sigma\tau}(\bm{\epsilon}_{\mb{q}\sigma}\cdot\hat{\bm{f}}\uq)$ and $h_{\mb{q}1\tau}=w\uqt-y\uqt$ and $h_{\mb{q}2\tau}=x\uqt-z\uqt$ with the Bogoliubov coefficients given in the main text around Eq.~\eqref{eq:bogcoef}, and with $\bm{f}\uq$ given in Eq.~\eqref{eq:f}. It follows that we can write
	\begin{equation}\label{eq:chibetbet}
	\chi_{\mb{q}\tau,-\mb{q}'\tau'}^{\beta\beta}=\frac{A\uq A_{\mb{q}'}}{\nu\uqt\nu_{\mb{q}'\tau'}}\sum_{\sigma\sigma'}h_{\mb{q}\sigma\tau}h_{\mb{q}'\sigma'\tau'}\sum_{ij}\epsilon_{\mb{q}\sigma i}\epsilon_{\mb{q}'\sigma' j}\chi_{\mb{q}i,-\mb{q}'j}^{ff},
	\end{equation}
	where
	\begin{equation}
	\chi_{\mb{q}i,-\mb{q}'j}^{ff}=-2\sum_{n\neq 0}\frac{\braket{\psi_0|\hat{f}_{\mb{q}i}|\psi_n}\braket{\psi_n|\hat{f}_{-\mb{q}'j}|\psi_0}}{\varepsilon_n-\varepsilon_0}.
	\end{equation}
	If the response function is translationally invariant, that is, if $\chi_{ij}^{ff}(\mb{x},\mb{x}')=\chi_{ij}^{ff}(\mb{x}-\mb{x}')$, then from Eq.~\eqref{eq:chiFT} it follows that $\chi_{\mb{q}i,-\mb{q}'j}^{ff}=\chi_{\mb{q}i,-\mb{q}j}^{ff}\delta_{\mb{q}\mb{q}'}$, where
	\begin{equation}
	\chi_{\mb{q}i,-\mb{q}j}^{ff}=\frac{1}{V}\int\text{d}^3r\ \chi^{ff}_{ij}(\mb{r})\rme^{-i\mb{q}\cdot\mb{r}}.
	\end{equation}
	As we prove in Appendix~\ref{app:rotation}, if $\hat{R}_\theta\hat{H}_m\hat{R}_{\theta}^\dagger=\hat{H}_m$ where $\hat{R}_\theta$ is the unitary representation of an SO$(3)$ rotation around the vector $\mb{q}$, i.e. within the transverse plane spanned by $\bm{\epsilon}_{\mb{q}1}-\bm{\epsilon}_{\mb{q}2}$, then in gauges for which $\hat{R}_\theta\hat{\bm{f}}\x\hat{R}_\theta^\dagger=\hat{\bm{f}}'(\mb{x})$, one may write that
	\begin{equation}\label{eq:rotinvar}
	\sum_{ij}\epsilon_{\mb{q}\sigma i}\epsilon_{\mb{q}\sigma' j}\chi_{\mb{q}i,-\mb{q}j}^{ff}=\chi_{\text{T}\mb{q}}^{ff}\delta_{\sigma\sigma'},
	\end{equation}
where `$\text{T}$' is the transverse component with respect to $\mb{q}$. We prove in Appendix~\ref{app:rotation} that the Coulomb and multipolar gauges meet this condition. Substitution of these results into Eq.~(\ref{eq:chibetbet}) yields
	\begin{equation}\label{eq:chibetbet2}
	\chi_{\mb{q}\tau,-\mb{q}'\tau'}^{\beta\beta}=\frac{A\uq^2}{\nu\uqt^2}\Lambda_{\mb{q}\tau\tau'}\chi_{\text{T}\mb{q}}^{ff}\delta_{\mb{q}\mb{q}'},
	\end{equation}
	where $\Lambda_{\mb{q}\tau\tau'}=\sum_\sigma h_{\mb{q}\sigma\tau}h_{\mb{q}\sigma\tau'}$. Substituting Eq.~(\ref{eq:chibetbet2}) into Eq.~\eqref{eq:cond} finally yields
	\begin{equation}\label{eq:Ft}
	\beta\uqt-\beta_{\mb{q}\tau 0}=V\frac{A\uq^2}{\nu\uqt^2}\chi^{ff}_{\text{T}\mb{q}}\sum_{\tau'}\Lambda_{\mb{q}\tau\tau'}F_{\mb{q}\tau'}.
	\end{equation}

	To solve for $F\uqt$ we must remove the remaining summation over $\tau'$, which requires $\Lambda_{\mb{q}\tau\tau'}\propto\delta_{\tau\tau'}$. Using the definitions of the Bogoliubov coefficients given in Appendix~\ref{app:diag} around Eq.~\eqref{eq:bogcoef} we find that,
	\begin{align}
	\Lambda_{\mb{q}\tau\tau'}=\left(1+\Phi\uqt\Phi_{\mb{q}\tau'}\right)\frac{\left(1+\Theta\uqt\right)\left(1+\Theta_{\mb{q}\tau'}\right)}{\sqrt{N\uqt N_{\mb{q}\tau'}}}.
	\end{align}
	In order that $\Lambda_{\mb{q}\tau\tau'}\propto\delta_{\tau\tau'}$ we require that $\Phi_{\mb{q}\pm}=\pm 1$. Recalling that $\Phi\uqpm=d\uq \pm (1+d\uq^2)^{1/2}$ where $d\uq=(D_{\mb{q}11}-D_{\mb{q}22})/(2D_{\mb{q}12})$ we see that $\Phi_{\mb{q}\pm}=\pm 1$ if and only if $d\uq=0$, which is satisfied if either $D_{\mb{q}11}=D_{\mb{q}22}$ or $D_{\mb{q}11}-D_{\mb{q}22}\ll 2D_{\mb{q}12}$. In this case $N\uqt=8\lambda\uqt(1-\lambda\uqt)^{-2}$ and therefore
	\begin{align}
		\lim_{d\uq\to0}\Lambda_{\mb{q}\tau\tau'}=\frac{\delta_{\tau\tau'}}{\lambda\uqt}.
	\end{align}
	Substituting this expression into Eq.~\eqref{eq:Ft} we find that $VF_{-\mb{q}\tau}=(\delta\beta\uqt^*\nu\uqt^2\lambda\uqt)/(A\uq^2\chi_{\text{T}\mb{q}}^{ff})$. Finally, using this expression for $F_{-\mb{q}\tau}$ in Eq.~\eqref{eq:stiff2} yields the result used in the main text:
	\begin{equation}
		E(\beta)=E(0)-\frac{1}{2}\sum_{\mb{q}\tau}\frac{\nu\uqt^2\lambda\uqt}{A\uq^2\chi_{\text{T}\mb{q}}^{ff}}\left|\beta_{\mb{q}\tau}-\beta_{\mb{q}\tau 0}\right|^2.
	\end{equation}

	\section{Translational invariance of the static linear response functions}\label{app:TIchi}

	In this appendix we prove that translational invariance of $\hat{H}_m$ implies that the static linear response function $\chi^{ff}_{\mb{q}i,-\mb{q}'j}$ encountered throughout the main text is also translationally invariant, and so $\chi^{ff}_{\mb{q}i,-\mb{q}'j}=\chi^{ff}_{\mb{q}i,-\mb{q}j}\delta_{\mb{q}\mb{q}'}$.
	
	In position space, the static linear response function is 
	\begin{equation}\label{eq:chiffapp}
		\chi^{ff}_{ij}(\mb{x},\mb{x}')=\left\langle\left[\hat{f}_i\x,\hat{f}_j\xp\right]\right\rangle_0,
	\end{equation}
where $\langle\cdot\rangle_0$ indicates the expectation value with respect to the ground matter state, $\ket{\psi_m^0}$. We will prove that translational invariance of $\hat{H}_m$ (and so of $\ket{\psi_m^0}$) immediately gives translational invariance of the response function, such that $\chi_{ij}^{ff}(\mb{x},\mb{x}')=\chi_{ij}^{ff}(\mb{x}-\mb{x}')$. In Appendix~\ref{app:TIHm}, we proved that all eigenstates of $\hat{H}_m$ are translationally invariant. Therefore, the proof in this appendix trivially extends to finite temperature.

The coupling operator in the response function is $\hat{f}\x=\sum\uq \hat{f}\uq\rme^{i\mb{q}\cdot\mb{x}}$ where
\begin{equation}\label{eq:fapp}
	\hat{\bm{f}}_{\mb{q}}= iV\nu\uq\left[\left(\check{\mb{q}}\times\hat{\mb{M}}^p_{\mb{q}}\right)+\hat{\mb{P}}_{\text{T}\mb{q}}\right],
\end{equation}
and $\hat{\mb{M}}^p\uq$ and $\hat{\mb{P}}_{\text{T}\mb{q}}$ are the Fourier components of the paramagnetisation and transverse polarisation. In Appendix~\ref{app:origin}, we have already shown that the polarisation is translated $\hat{T}(a)$ [defined in Eq.~\eqref{eq:trans2}] such that in position space, $\hat{T}({\bf a})\hat{{\bf P}}_{\rm T}({\bf x})\hat{T}({\bf a})^\dagger =\hat{{\bf P}}_{\rm T}({\bf x}-{\bf a})$. We must now prove that the paramagnetisation translates in the same way.

The paramagnetisation can be written in terms of $\mb{g}_\text{T}=\mb{0}$ and $\mb{g}_\text{T}$-dependent contributions as $\hat{{\bf M}}^p\x=\hat{{\bf M}}_0^p\x+\hat{{\bf M}}_{\gee_{\rm T}}^p\x$ which are given in Eqs.~\eqref{eq:MV0} and \eqref{eq:MVg}. Note that if ${\bf y} = {\bf x}-{\bf a}$ then $\partial/\partial y_i = \sum_j (\partial x_j/ \partial y_i) (\partial / \partial x_j) = \partial/\partial x_i $. 
Consider first the $\mb{g}_\text{T}=\mb{0}$ contribution,
\begin{align}
	\hat{T}({\bf a})\hat{{\bf M}}_0^p({\bf x})\hat{T}({\bf a})^\dagger &= \int d^3 x' {\nabla_{\mb{x}'} \times \hat{{\bf j}}^p({\bf x}'-{\bf a}) \over 4\pi|{\bf x'}-{\bf x}|} \nonumber\\
	&= \int d^3 x' {\nabla_{{\bf x}'-{\bf a}} \times \hat{{\bf j}}^p({\bf x}'-{\bf a}) \over 4\pi|{\bf x'}-{\bf x}|} \nonumber\\
	&=  \int d^3 x' {\nabla_{\mb{x}'} \times \hat{{\bf j}}^p({\bf x}') \over 4\pi|{\bf x'}-({\bf x}-{\bf a})|} = \hat{{\bf M}}_0^p({\bf x}-{\bf a}).
\end{align}
The proof for the $\mb{g}_\text{T}$-dependent contribution follows similarly. In the case that ${\bf g}_{\rm T}({\bf x},{\bf x}')={\bf g}_{\rm T}({\bf x}-{\bf x}')$, we have
\begin{align}
	\hat{T}({\bf a})\hat{{\bf M}}_{{\gee}_{\rm T}}^p({\bf x})\hat{T}({\bf a})^\dagger &= \int d^3 x' \hat{{\bf j}}^p({\bf x}'-{\bf a}) \cdot \nabla_{\mb{x}'} \int d^3y{\nabla_{\bf y} \times {\bf g}_{\rm T}({\bf y}-{\bf x}') \over 4\pi|{\bf y}-{\bf x}|} \nonumber\\
	&= \int d^3 x' \hat{{\bf j}}^p({\bf x}') \cdot \nabla_{\mb{x}'} \int d^3y{\nabla_{\bf y} \times {\bf g}_{\rm T}({\bf y}-{\bf a}-{\bf x}') \over 4\pi|{\bf y}-{\bf x}|} \nonumber \\ &= \int d^3 x' \hat{{\bf j}}^p({\bf x}') \cdot \nabla_{\mb{x}'} \int d^3y{\nabla_{\bf y} \times {\bf g}_{\rm T}({\bf y}-{\bf x'}) \over 4\pi|{\bf y}-({\bf x}-{\bf a})|} =\hat{{\bf M}}_{{\gee}_{\rm T}}^p({\bf x}-{\bf a}).
\end{align}
For the transverse Green's function of the multipolar-gauge, given in Eq.~(\ref{go}), we have
\begin{align}
	\hat{T}({\bf a})\hat{{\bf M}}_{{\gee}_{\rm T}}^p({\bf x})\hat{T}({\bf a})^\dagger &= \int d^3 x' \hat{{\bf j}}^p({\bf x}'-{\bf a}) \cdot \nabla_{\mb{x}'} \int d^3y{\nabla_{\bf y} \times {\bf g}_{\rm T}({\bf y}-{\bf a},{\bf x}'-{\bf a}) \over 4\pi|{\bf y}-{\bf x}|}\nonumber\\
	& = \int d^3 x' \hat{{\bf j}}^p({\bf x}') \cdot \nabla_{\mb{x}'} \int d^3y{\nabla_{\bf y} \times {\bf g}_{\rm T}({\bf y},{\bf x}') \over 4\pi|{\bf y}-({\bf x}-{\bf a})|} \nonumber \\ &= \hat{{\bf M}}_{{\gee}_{\rm T}}^p({\bf x}-{\bf a}).
\end{align}
Therefore the full paramagnetisation is translationally invariant in all cases considered in the main text. That is, $\hat{T}({\bf a})\hat{{\bf M}}^p({\bf x})\hat{T}({\bf a})^\dagger = \hat{ {\bf M}}^p({\bf x}-{\bf a})$.

We have now proven that both the transverse polarisation and the paramagnetisation translate as $\hat{T}({\bf a})\hat{F}({\bf x})\hat{T}({\bf a})^\dagger =  \hat{F}({\bf x}-{\bf a})$. In momentum space this translation becomes
\begin{align}
	\hat{T}({\bf a})\hat{F}_{\bf q}\hat{T}({\bf a})^\dagger  = {1\over V}\int d^3 x  \hat{F}({\bf x}-{\bf a})e^{-i{\bf q}\cdot {\bf x}} =  {1\over V}\int d^3 x  F({\bf x})e^{-i{\bf q}\cdot {\bf x}} e^{-i{\bf q}\cdot {\bf a}} = \hat{F}_{\bf q}e^{-i{\bf q}\cdot {\bf a}}.
\end{align}
Applying this equality to $\hat{{\bf P}}_{\rm T\mb{q}}$ and $\hat{{\bf M}}^p\uq$ in Eq.~\eqref{eq:fapp} we find that
\begin{align}
	\hat{T}({\bf a})\hat{{\bm f}}_{\bf q}\hat{T}({\bf a})^\dagger  = \hat{{\bm f}}_{\bf q}e^{-i{\bf q}\cdot {\bf a}}.
\end{align}
Inverting the Fourier transform then gives
\begin{align}
	\hat{T}({\bf a})\hat{{\bm f}}({\bf x})\hat{T}({\bf a})^\dagger = \sum_{\bf q} \hat{{\bm f}}_{\bf q}e^{-i{\bf q}\cdot {\bf a}}e^{i{\bf q}\cdot {\bf x}} = \hat{{\bm f}}({\bf x}-{\bf a}).
\end{align}

We are now in a position to prove that the static linear response function in Eq.~\eqref{eq:chiffapp} is translationally invariant,
\begin{align}
	\chi_{ij}^{ff}({\bf x},{\bf x}') = \bra{\psi_m^0} [\hat{f}_i({\bf x}), \hat{f}_j({\bf x}')] \ket{\psi_m^0}= \bra{\psi_m^0}\hat{T}({\bf a})^\dagger [\hat{f}_i({\bf x}), \hat{f}_j({\bf x}')] \hat{T}({\bf a})\ket{\psi_m^0} = \bra{\psi_m^0} [\hat{f}_i({\bf x}+{\bf a}), \hat{f}_j({\bf x}'+{\bf a})]\ket{\psi_m^0}. 
\end{align}
Choosing ${\bf a}=-{\bf x}'$ we obtain $\chi_{ij}^{ff}({\bf x},{\bf x}') = \chi_{ij}^{ff}({\bf x}-{\bf x}',{\bf 0})$ which completes the proof.

\section{Static linear response function and rotations}\label{app:rotation}

In this appendix we consider the static linear response function
\begin{equation}\label{eq:appchiff}
	\tilde{\chi}_{\mb{q},-\mb{q}}^{{f}{f}}=-2V\sum_{n\neq 0}\frac{\braket{\psi_m^0|\hat{\bm f}_{\mb{q}}|\psi_m^n}\braket{\psi_m^n|\hat{\bm f}_{-\mb{q}}|\psi_m^0}}{\varepsilon_m^{(n)}-\varepsilon_m^{(0)}},
\end{equation}
where the juxtaposition of vectors on the numerator denotes the tensor product of the vector-valued matrix elements in Euclidean three space. Eq.~\eqref{eq:appchiff} relates to the relevant static linear response function in the main text by $\tilde{\chi}^{ff}_{\mb{q}\sigma,-\mb{q}\sigma'}=\bm{\epsilon}\uqs\cdot\tilde{\chi}_{\mb{q},-\mb{q}}^{{f}{f}}\cdot\bm{\epsilon}_{\mb{q}\sigma'}$.

We denote by $\hat{R}_\theta$ the unitary representation of an ${\rm SO}(3)$ rotation $S_\theta$, which rotates the canonical operators of each charge through an angle $\theta$, and we denote the rotated operators with a prime, i.e., 
$\hat{\mb{p}}_\mu':=S_\theta {\hat {\bf p}}_\mu = \hat{R}_\theta\hat{\mb{p}}_\mu \hat{R}_\theta^\dagger$ and $\hat{\mb{r}}_\mu'=S_\theta {\hat {\bf r}}_\mu =\hat{R}_\theta\hat{\mb{r}}_\mu \hat{R}_\theta^\dagger$. The generator of rotations is the total angular momentum operator, which commutes with the Hamiltonian. It follows that ${\hat R}_\theta {\dot {\hat {\bf r}}}_\mu {\hat R}_\theta^\dagger = {\dot {\hat {\bf r}}}_\mu'$ and therefore that ${\hat R}_\theta {\hat {\bf j}}({\bf x}){\hat R}_\theta^\dagger = S_\theta {\hat {\bf j}}(S_\theta^T{\bf x})$. In Fourier space this transformation property of the current reads
${\hat R}_\theta {\hat {\bf j}}_{\bf q}{\hat R}_\theta^\dagger = S_\theta {\hat {\bf j}}_{S_\theta^T{\bf q}}$. We assume that the polarisation field $\hat{{\bf P}}$ possesses the same transformation property, noting that this is indeed the case in, for example, the Coulomb and multipolar gauges. In the latter case the polarisation reads
\begin{equation}\label{eq:PTqmultiolar}
	\hat{{\bf P}}({\bf q}) = -{e\over V} \left[\sum_\mu {\hat{{\bf r}}_\mu -\hat{{\bf o}}\over i{\bf q}\cdot( \hat{\bf r}_\mu-\hat{\bf o})}[e^{i{\bf q}\cdot \hat{\bf r}_\mu}-e^{i{\bf q}\cdot \hat{\bf o}} ]-{N\over V}\int d^3 x' {{\bf x}' -\hat{{\bf o}}\over i{\bf q}\cdot ({\bf x}'-\hat{{\bf o}})}[e^{i{\bf q}\cdot {\bf x}'}-e^{i{\bf q}\cdot \hat{\bf o}} ]\right].
\end{equation}
We note further that since both $\hat{{\bf j}}$ and $\hat{{\bf P}}$ possess the same transformation property it follows that $\nabla \times \hat{{\bf M}}\x$ possesses this transformation property as well.  

The transformation property of the polarisation implies that the matter Hamiltonian $\hat{H}_m$ is rotationally invariant, that is,
\begin{equation}
	{\hat R}_\theta\hat{H}_m {\hat R}_\theta^\dagger = \sum_\mu {({\hat {\bf p}}'_\mu)^2 \over 2m} +\frac{1}{2}\int d^3 x \hat{{\bf P}}'({\bf x}')^2 =  \sum_\mu {{\hat {\bf p}}_\mu^2 \over 2m} +\frac{1}{2}\int d^3 x' \hat{{\bf P}}({\bf x}')^2 = \hat{H}_m
\end{equation}
where ${\bf x'} = S_\theta^T{\bf x}$, and we have used $d^3x =d^3x'$ and ${\bf a}'\cdot {\bf b}'={\bf a}\cdot {\bf b}$ for arbitrary vectors ${\bf a}$ and ${\bf b}$. It follows that  $\hat{R}_\theta^\dagger \ket{\psi_m^n}=\rme^{i\theta_n}\ket{\psi_m^n}$ where $\theta_n$ is real. 

We now restrict our attention to rotations around the vector ${\bf q}$, i.e., in the transverse plane spanned by the ${\bm \epsilon}_{{\bf q}\sigma},\,\sigma=1,2$. For notational economy, we suppress the index $_{\bf q}$ on these polarisation vectors for the remainder of this Appendix. The rotation leaves ${\bf q}$ invariant but it rotates the orthogonal polarisation vectors. The paramagnetic current ${\hat {\bf j}}^p$ possesses the same transformation property as ${\hat {\bf j}}$ and, under a rotation around ${\bf q}$, so does the transverse polarisation ${\hat {\bf P}}_{\rm T}$. Therefore, the field ${\mb{q}}\times\hat{\mb{M}}^p_{\mb{q}}$ must also transform as ${\hat R}_\theta {\mb{q}}\times\hat{\mb{M}}^p_{\mb{q}}{\hat R}_\theta^\dagger ={\mb{q}}\times\hat{\mb{M}}'^p_{\mb{q}}$ where $\hat{{\bf M}}'^p=S_\theta \hat{{\bf M}}^p$ and $S_\theta$ refers to a rotation in the ${\bm \epsilon}_1$-${\bm \epsilon}_2$ plane. It follows that 
\begin{equation}\label{eq;tr}
	{\hat R}_\theta \hat{\bm{f}}_{\mb{q}}{\hat R}_\theta^\dagger = iV\nu\uq\left[\left(\check{\mb{q}}\times\hat{\mb{M}}'^p_{\mb{q}}\right)+\hat{\mb{P}}'_{\text{T}\mb{q}}\right] = {\hat {\bm f}}'_{\bf q}
\end{equation}

Since $\hat{R}_\theta^\dagger \ket{\psi_m^n}=\rme^{i\theta_n}\ket{\psi_m^n}$ the static linear response function in Eq.~\eqref{eq:appchiff} can be written
\begin{equation}\label{eq:appchiff2}
	\tilde{\chi}_{\mb{q},-\mb{q}}^{{f}{f}}=-2V\sum_{n\neq 0}\frac{\braket{\psi_m^0|{\hat{\bm f}}'_{\mb{q}} |\psi_m^n}\braket{\psi_m^n|{\hat{\bm f}}'_{-\mb{q}}|\psi_m^0}}{\varepsilon_m^{(n)}-\varepsilon_m^{(0)}} = \tilde{\chi}_{\mb{q},-\mb{q}}^{{f}'{f}'}.
\end{equation}
The components of this response function in the ${\bm \epsilon}_1$-${\bm \epsilon}_2$ plane are
\begin{equation}\label{eq:appchiff3}
	\tilde{\chi}_{\mb{q}\sigma,-\mb{q}\sigma'}^{ff}:= {\bm \epsilon}_\sigma \cdot \tilde{\chi}_{\mb{q},-\mb{q}}^{{f}{f}} \cdot {\bm \epsilon}_{\sigma'}= {\bm \epsilon}_\sigma \cdot \tilde{\chi}_{\mb{q},-\mb{q}}^{{f}'{f}'} \cdot {\bm \epsilon}_{\sigma'}
\end{equation}
where the second equality follows from Eq.~(\ref{eq:appchiff2}). Consider now a rotation by $\pi/2$ such that $S_\theta {\bm \epsilon}_1 =  -{\bm \epsilon}_2$ and $S_\theta {\bm \epsilon}_2  = {\bm \epsilon}_1$, so that ${\bm f}'^{0n}_{\bf q} = S_{\pi/2}{\bm f}_{\bf q}^{0n} = -f_{{\bf q}1}^{0n}{\bm \epsilon}_2 + f_{{\bf q}2}^{0n}{\bm \epsilon}_1$ where ${\bm f}_{\mb{q}\sigma}^{0n} =\braket{\psi_m^0|\hat{f}_{\mb{q}\sigma} |\psi_m^n}$ and $f^{0n}_{{\bf q}\sigma} :={\bm \epsilon}_\sigma \cdot {\bm f}^{0n}_{\bf q}$. Substitution of this expression into Eq.~(\ref{eq:appchiff3}) implies that $\tilde{\chi}_{{\bf q}1,-{\bf q}1}^{ff} = \tilde{\chi}_{{\bf q}2,-{\bf q}2}^{ff}=:\tilde{\chi}^{ff}_{{\rm T}{\bf q}}$ and $\tilde{\chi}_{{\bf q}1,-{\bf q}2}^{ff} = -\tilde{\chi}_{{\bf q}2,-{\bf q}1}^{ff}$. If it is also the case that $\tilde{\chi}_{\mb{q}\sigma,-\mb{q}\sigma'}^{ff}$ is symmetric, i.e., $\tilde{\chi}_{\mb{q}\sigma,-\mb{q}\sigma'}^{ff}=\tilde{\chi}_{\mb{q}\sigma',-\mb{q}\sigma}^{ff}$ for $\sigma\neq \sigma'$, then it follows that
\begin{align}
	\sum_{ij}\epsilon_{\mb{q}\sigma i}\epsilon_{\mb{q}\sigma' j}\tilde{\chi}_{\mb{q}i,-\mb{q}j}^{ff}\equiv \tilde{\chi}_{\mb{q}\sigma,-\mb{q}\sigma'}^{ff} = \tilde{\chi}^{ff}_{\rm T{\bf q}}\delta_{\sigma\sigma'}
\end{align}
which is Eq.~\eqref{eq:rotinvar}. We note that $\tilde{\chi}_{\mb{q}\sigma,-\mb{q}\sigma'}^{ff}$ is symmetric in, for example, the Coulomb and dipole gauges where the symmetry of $\tilde{\chi}_{\mb{q}\sigma,-\mb{q}\sigma'}^{ff}$ is equivalent to the reality of $\tilde{\chi}_{\mb{q}\sigma,-\mb{q}\sigma'}^{ff}$.

	\section{Derivation of the Fourier response amplitudes}\label{app:LinearResponse}
	In this appendix we prove that the average change of an operator $\hat{O}_i$  due to a perturbation away from $\hat{H}_\text{eq}$ in the form $\intx \hat{\mb{C}}\x\cdot \mb{F}(\mb{x})$, where $\hat{\mb{C}}\x$ is some coupling operator and $\mb{F}(\mb{x})$ is the perturbing field, is $\langle\hat{O}_i\rangle_\delta=\sum\uq O_{\mb{q}i}^\delta\rme^{i\mb{q}\cdot\mb{x}}$ where
	\begin{equation}\label{eq:respamp}
		O^{\delta}_{\mb{q}i}=\sum_{ j}\tilde{\chi}^{OC}_{\mb{q}i,-\mb{q}j} F_{\mb{q}j}.
	\end{equation}
	We will use this result in subsequent appendices. According to linear response theory (see, for example, Section 3.2 of Ref.~\cite{giuliani2005quantum}) the perturbation of $\langle \hat{O}_i\x\rangle$ away from its thermal equilibrium value, $\langle \hat{O}_i\x\rangle_\text{eq}$, defined using Hamiltonian $\hat{H}_\text{eq}$, is to first order in $\mb{F}\x$ given by
	\begin{align}\label{eq:Oresp}
		\langle \hat{O}_i\x\rangle_\delta&=\langle \hat{O}_i\x\rangle-\langle\hat{O}_i\x\rangle_\text{eq}\\
		&=\sum_j\int_0^\infty\text{d}\tau\ \intxp \zeta^{OC}_{ij}(\mb{x},\mb{x}',\tau)F_j(\mb{x}',t-\tau),\nonumber
	\end{align}
	where
	\begin{equation}
		\zeta^{OC}_{ij}(\mb{x},\mb{x}',\tau)=-i\left\langle\left[\hat{O}_{i}(\mb{x},\tau),\hat{C}_{j}\xp\right]\right\rangle_\text{eq},
	\end{equation}
	is a linear response function. The brackets $\langle\cdot\rangle_\text{eq}$ denote the expectation value with respect to the thermal equilibrium Gibbs state defined by $\hat{H}_\text{eq}$, which at zero temperature, is just the ground state of $\hat{H}_\text{eq}$. Assuming that the perturbing field is static $\mb{F}(\mb{x},t)=\mb{F}\x$, the time integral yields $\int\text{d}\tau\ \zeta^{OC}_{ij}(\mb{x},\mb{x}',\tau)= \zeta^{OC}_{ij}(\mb{x},\mb{x}',\omega=0)\equiv \chi_{ij}^{OC}(\mb{x},\mb{x}')$. Note that we have reserved $\chi$ to denote \textit{static} linear response functions, i.e. evaluated at $\omega=0$. The amplitude of the response along direction $\check{\mb{q}}$, denoted $O_{\mb{q}i}^\delta$ where $\langle \hat{O}_i\x\rangle_\delta=\sum\uq O_{\mb{q}i}^\delta\rme^{i\mb{q}\cdot\mb{x}}$, is found via Fourier transformation of the right-hand-side of Eq.~\eqref{eq:Oresp} after assuming a static perturbing field. We find that
	\begin{equation}
		O^{\delta}_{\mb{q}i}=V\sum_{\mb{q}' j}\chi^{OC}_{i\mb{q},-\mb{q}'j} F_{\mb{q}' j},
	\end{equation}
	where
	\begin{align}
		\chi^{OC}_{i\mb{q},-\mb{q}'j}=\frac{1}{V^2}&\intx \rme^{-i\mb{q}\cdot\mb{x}}\intxp\rme^{i\mb{q}'\cdot\mb{x}'}\chi^{OC}_{ij}(\mb{x},\mb{x}').
	\end{align}
	Translational invariance of the static linear response function, $\chi_{ij}^{OC}(\mb{x},\mb{x}')=\chi^{OC}_{ij}(\mb{x}-\mb{x}')$, implies $\chi^{OC}_{i\mb{q},-\mb{q}'j}=\chi^{OC}_{\mb{q}i,-\mb{q}j}\delta_{\mb{q},\mb{q}'}$, where 
	\begin{equation}\label{eq:chiO}
		\chi^{OC}_{\mb{q}i,-\mb{q}j}=\frac{1}{V}\int\text{d}^3r\ \chi^{OC}_{ij}(\mb{r})\rme^{-i\mb{q}\cdot\mb{r}}.
	\end{equation}
	It follows that the linear response amplitude is given by Eq.~(\ref{eq:respamp}) in which $\tilde{\chi}^{OC}_{\mb{q}i,-\mb{q}j}=V\chi^{OC}_{\mb{q}i,-\mb{q}j}$. This completes the proof. 
	We note that after expansion into the matter eigenbasis at zero temperature, one can show that Eq.~\eqref{eq:chiO} is equivalent to the Lehmann representation in Eq.~\eqref{eq:chiAB} \cite{giuliani2005quantum}.

	\section{Magnetic response to the magnetic interaction in the Coulomb gauge\label{app:mag}}
	In this Appendix we derive the total Coulomb gauge magnetisation-magnetisation static linear response function, $\tilde{\chi}^{MM}_{\mb{q}i,-\mb{q}j}$, appearing in the main text. This is the linear response of the total Coulomb gauge magnetisation operator, $\hat{\mb{M}}\x=\hat{\mb{M}}^p\x+\hat{\mb{M}}^d\x$, to the total magnetic perturbation,
	\begin{equation}\label{eq:HBp}
		\hat{H}_B'=-\intx\hat{\mb{M}}\x\cdot\mb{B}\x.
	\end{equation}
	Note that as usual in linear response theory [see Appendix~\ref{app:LinearResponse}] the perturbing field is treated classically. 
	
	The Coulomb gauge is defined by $\mb{g}_\text{T}=\mb{0}$ which, recalling from Appendix~\ref{app:HB}, means that the total magnetisation is $\hat{\mb{M}}\uq\x=\hat{\mb{M}}_{\mb{q}}^p\x+\hat{\mb{M}}_{\mb{q}}^d\x$ where
	\begin{align}
		\hat{\mb{M}}^\xi\x=\intxp\frac{\bm{\nabla}_{\mb{x}'}\times\hat{\mb{j}}^\xi(\mb{x}')}{4\pi\left|\mb{x}-\mb{x}'\right|},\label{eq:M0}
	\end{align} 
	and $\xi=p,d$ labels the para- and dia-magnetisations. The subscript `$0$' labels the $\mb{g}_\text{T}=\mb{0}$ dependent part, which is the only contribution in the Coulomb-gauge. The para- and dia-magnetic current densities are
	\begin{align}
		\hat{\mb{j}}^p\x&=-\frac{e}{2m}\sum_\mu\{\hat{\mb{p}}_\mu,\delta(\mb{x}-\hat{\mb{r}}_\mu)\},\label{eq:jpapp}\\
		\hat{\mb{j}}^d\x&=-\frac{e^2}{m}\mb{A}\x \hat{n}_e\x,\label{eq:jdapp}
	\end{align} 
	which sum to the total, gauge-invariant current density, $\hat{\mb{j}}\x=\hat{\mb{j}}^p\x+\hat{\mb{j}}^d\x=-(e/2)\sum_\mu\{\dot{\hat{\mb{r}}}_\mu,\delta(\mb{x}-\hat{\mb{r}}_\mu)\}$. 
	
	Using the notation we have introduced for linear response theory in Appendix~\ref{app:LinearResponse}, the static linear response of the $i^\text{th}$ component of $\langle\hat{\mb{M}}\x\rangle$ away from equilibrium, due to $\hat{H}_B'$ in Eq.~\eqref{eq:HBp}, is given by
	\begin{equation}\label{eq:ChiMMInitial}
		\left\langle\hat{M}_{i}\x\right\rangle - \left\langle\hat{M}_{i}\x\right\rangle_\text{eq}= -\sum_j\intxp\chi^{MM}_{ij}(\mb{x},\mb{x}')B_j\xp,
	\end{equation}
	where the position space static linear response function is
	\begin{equation}
		\chi^{MM}_{ij}(\mb{x},\mb{x}')=-i\left\langle\left[\hat{M}_{i}(\mb{x}),\hat{M}_{j}\xp\right]\right\rangle_\text{eq}.
	\end{equation}
	
	We now make two remarks about the para- and dia-magnetisation to simplify Eq.~\eqref{eq:ChiMMInitial}. 
	\begin{enumerate}
		\item Since $\hat{\mb{j}}^d\x$ [Eq.~\eqref{eq:jdapp}] depends on $\mb{A}\x$ it follows that $\hat{\mb{M}}^d\x$ is linearly dependent on the amplitude of the magnetic field. This has two immediate implications: (a) $\langle\hat{M}^d_{ i}\x\rangle_\text{eq}$ does not contribute to the equilibrium [$\mb{B}\x=\mb{0}$] value of the total magnetisation operator and instead contributes to the linear response away from equilibrium; (b) to zeroth order in the perturbing field, $\mb{B}\x$, the static linear response function in Eq.~\eqref{eq:ChiMMInitial} must be
		\begin{equation}
			\chi^{MM}_{ij}(\mb{x},\mb{x}')\approx\chi^{M^pM^p}_{ij}(\mb{x},\mb{x}'),
		\end{equation}
		to ensures that the right-hand-side of Eq.~\eqref{eq:ChiMMInitial} is first order in the perturbing field.
		
		\item Unlike the diamagnetic current, the paramagnetic current in Eq.~\eqref{eq:jpapp} is not proportional to $\mb{A}\x$. Therefore, it is clear that $\langle\hat{M}^p_{ i}\rangle_\text{eq}$ will not be linear in the perturbing magnetic field and so it will contribute to the equilibrium value.  This fact, along with point (1), also implies that in the Coulomb-gauge $\langle\hat{M}^p_{i}\rangle_\text{eq}$ entirely determines the equilibrium value of the total magnetisation.
	\end{enumerate}
	
	It follows from points (1) and (2) that we can rewrite Eq.~\eqref{eq:ChiMMInitial} as 
	\begin{equation}\label{eq:ChiMMInitial2} 
		\left\langle\hat{M}_{i}\x\right\rangle - \left\langle\hat{M}^p_{i}\x\right\rangle_\text{eq}=\left\langle\hat{M}^d_{i}\x\right\rangle_\text{eq}- \sum_j\intxp\chi^{M^pM^p}_{ij}(\mb{x},\mb{x}')B_j\xp,
	\end{equation}
	where terms on the right-hand-side define the linear response. Taking the Fourier transformation of the right-hand-side of Eq.~\eqref{eq:ChiMMInitial2} yields
	\begin{equation}
		\langle \hat{M}_{i}\x\rangle-\langle \hat{M}^p_{i}\x\rangle_\text{eq}=\sum\uq M^\delta_{\mb{q}i}\rme^{i\mb{q}\cdot\mb{x}},
	\end{equation}
	where we have defined the Fourier amplitude,
	\begin{equation}\label{eq:ChiMMAmp}
		M^\delta_{\mb{q}i}=\langle \hat{M}^d_{\mb{q}i}\rangle_\text{eq}-\sum_j\tilde{\chi}_{\mb{q}i,-\mb{q}j}^{M^pM^p}B_{\mb{q}j},
	\end{equation} 
	with $\langle \hat{M}^d_{i}\x\rangle_\text{eq}=\sum\uq \langle \hat{M}^d_{\mb{q}i}\rangle_\text{eq}\rme^{i\mb{q}\cdot\mb{x}}$ and
	\begin{equation}
		\chi^{M^pM^p}_{\mb{q}i,-\mb{q}j}=\frac{1}{V}\int\text{d}^3r\ \chi^{M^pM^p}_{ij}(\mb{r})\rme^{-i\mb{q}\cdot\mb{r}}.
	\end{equation} 
The response function $\tilde{\chi}_{\mb{q}i,-\mb{q}j}^{M^pM^p}=V\chi_{\mb{q}i,-\mb{q}j}^{M^pM^p}$ is dimensionless, and we have used the translational invariance of the paramagnetic response function, $\chi_{ij}^{M^pM^p}(\mb{x},\mb{x}')=\chi_{ij}^{M^pM^p}(\mb{x}-\mb{x}')\equiv\chi_{ij}^{M^pM^p}(\mb{r})$ (see Appendix~\ref{app:LinearResponse}).
	
	Finally we must obtain an expression for  $\langle \hat{M}^d_{\mb{q}i}\rangle_\text{eq}$ in terms of the perturbing field, $\mb{B}\x$. Taking the Fourier transforms of Eq.~\eqref{eq:M0} and Eq.~\eqref{eq:jdapp} yields
	\begin{equation}
		\hat{\mb{M}}^d_{\mb{q}}=i\frac{\check{\mb{q}}\times\hat{\mb{j}}^d\uq}{\nu\uq},
	\end{equation} 
	and $\hat{\mb{j}}^d\uq=-(e^2/m)\sum
	_\mb{k} \mb{A}_{\mb{q}-\mb{k}}\hat{n}_{e\mb{k}}$, respectively. At zero temperature, uniformly distributed charges in the ground state implies $\langle \hat{\mb{j}}^d\uq\rangle_\text{eq}=-[(e^2N)/(mV)]\mb{A}\uq$, such that the Fourier component of the equilibrium diamagnetisation is
	\begin{equation}\label{eq:ChiMdApp}
		\langle \hat{M}^d_{\mb{q}i}\rangle_\text{eq}=\tilde{\chi}_{\mb{q}}^{M^d}B_{\mb{q}i},
	\end{equation}
	where $\tilde{\chi}^{M^d}\uq=-(e^2N)/(mV\nu\uq^2)$ is the diamagnetic static linear response function as given in the main text and $\mb{B}\uq=i\nu\uq\check{\mb{q}}\times\mb{A}\uq$. Together, Eqs.~\eqref{eq:ChiMMAmp} and \eqref{eq:ChiMdApp} give the expression in the main text, namely, $M^\delta_{g\mb{q}i}=-\sum_j\tilde{\chi}_{\mb{q}i,-\mb{q}j}^{MM}B_{\mb{q}j}$ with
	\begin{equation}
		\tilde{\chi}_{\mb{q}i,-\mb{q}j}^{MM}=\tilde{\chi}_{\mb{q}i,-\mb{q}j}^{M^pM^p}-\delta_{ij}\tilde{\chi}_{\mb{q}}^{M^d}.
	\end{equation}

\section{Proof that the theory is gauge-invariant}\label{eq:ET}
In this appendix we calculate the expectation value of the transverse electric field, $\hat{\mb{E}}_\text{T}\x=-\hat{\bm{\Pi}}\x-\hat{\mb{P}}_\text{T}\x$ in the Coulomb- and dipole- gauges with respect to the separable light-matter state in the thermodynamic limit, $\ket{\Psi\{\beta\uqt\}}=\ket{\psi_m}\ket{\phi_l\{\beta\uqt\}}$. 
The dipole-gauge is the long wavelength limit of the multipolar-gauge, therefore to verify gauge-invariance we must also take the long wavelength limit in the Coulomb-gauge. We begin by reiterating the definitions of the photonic canonical momentum and transverse polarisation field,
\begin{align}
\hat{\bm{\Pi}}\x&=-i\sum_{\mb{q}\sigma}\nu\uq A\uq\bm{\epsilon}\uq\left(\rme^{i\mb{q}\cdot\mb{x}}\hat{a}\uq-\rme^{-i\mb{q}\cdot\mb{x}}\hat{a}\uq^\dagger\right),\label{eq:PiApp}\\
\hat{\mb{P}}_\text{T}\x&=-\intxp\mb{g}_\text{T}(\mb{x},\mb{x}')\hat{\rho}\xp.
\end{align}

We will write $\left\langle \cdot\right\rangle^g_s$ to denote the expectation value in gauge $g=C,d$ (Coulomb, dipole) with respect to state $s=m,l$ ($\ket{\psi_m},\ket{\phi_l\{\beta\uqt\}}$). If there is no $g$ label then $\mb{g}_\text{T}$ is unspecified and if  there is no $s$ label then the expectation value is taken with respect to the full light-matter state, $\ket{\Psi\{\beta\uqt\}}$. 

We begin by calculating the average photonic canonical momentum in an arbitrary gauge. Recall from the main text that the light state of the condensate is a coherent state with coherence $\left\langle \hat{c}\uqt \right\rangle_l\equiv\beta\uqt$ where $\hat{c}\uqt$ is the Bogoliubov transformed annihilation operator. The Bogoliubov operators are obtained by performing a Bogoliubov transformation to remove the diamagnetic interaction and are given by
\begin{align}
\hat{a}_{\mb{q} 1}&=w_{\mb{q}+}\hat{c}_{\mb{q}+}+w_{\mb{q}-}\hat{c}_{\mb{q}-}-y_{\mb{q}+}\hat{c}_{\mb{q}+}^\dagger-y_{\mb{q}-}\hat{c}_{\mb{q}-}^\dagger,\\
\hat{a}_{\mb{q} 2}&=x_{\mb{q}+}\hat{c}_{\mb{q}+}+x_{\mb{q}-}\hat{c}_{\mb{q}-}-z_{\mb{q}+}\hat{c}_{\mb{q}+}^\dagger-z_{\mb{q}-}\hat{c}_{\mb{q}-}^\dagger.
\end{align}
These definitions are the inversion of Eq.~\eqref{eq:bog}. The coefficients are given in their full form in Appendix~\ref{app:diag}, but here we note that the dipole- and Coulomb-gauges both have $d\uq=0$ where $d_{\bf q}$ is defined in Eq.~\eqref{eq:d}. It then follows that
\begin{align}\label{eq:wxyz}
y\uqt=-\tau\frac{\lambda\uqt-1}{2\sqrt{2\lambda\uqt}},\quad w\uqt=-\tau\frac{\lambda\uqt+1}{2\sqrt{2\lambda\uqt}},\quad
x\uqt=-\frac{\lambda\uqt+1}{2\sqrt{2\lambda\uqt}},\quad z\uqt=-\frac{\lambda\uqt-1}{2\sqrt{2\lambda\uqt}},
\end{align}
with $\tau\in\{+,-\}$. In the dipole- and Coulomb-gauges, $\lambda\uqt=1$ and $\lambda\uqt=\sqrt{1+[(e^2 N )/(m V\nu\uq^2)]}$, respectively. Note that in both gauges $\lambda\uqt=\lambda\uq$ but we will keep the $\tau$ label for generality.

The coherence $\beta_{{\bf q}\tau}$ is the expectation of $\hat{\beta}\uqt=-[A\uq/(\nu\uq\lambda\uqt)]\hat{g}\uqt$ [Eq.~\eqref{eq:betahat}], with respect to the matter state $\ket{\psi_m}$, where $\hat{g}\uqt=\sum_\sigma h_{\mb{q}\sigma\tau}\left(\bm{\epsilon}\uqs\cdot\hat{\bm{f}}\uq\right)$ is the coupling of the $\hat{c}\uqt$ mode to the cavity [with $h_{\mb{q}1\tau}=w\uqt-y\uqt$ and $h_{\mb{q}2\tau}=x\uqt-z\uqt$] and 
\begin{equation}
\hat{\bm{f}}\uq=i\nu\uq V\left[\check{\mb{q}}\times\hat{\mb{M}}^p\uq+\hat{\mb{P}}_{\text{T}\mb{q}}\right],
\end{equation}
is the coupling of the $\hat{a}\uqs$ mode to the cavity. Recall that $\check{\mb{q}} = {\bf q}/|{\bf q}|$. Due to the simpler form of Eqs.~\eqref{eq:wxyz}, the coupling coefficients within $\hat{g}\uqt$ also simplify to give $h_{\mb{q}1\tau}=-\tau/\sqrt{2\lambda\uqt}$ and $h_{\mb{q}2\tau}=-1/\sqrt{2\lambda\uqt}$. Substituting the definitions above into Eq.~\eqref{eq:PiApp}, one finds that
\begin{equation}\label{eq:PiExp}
\left\langle\hat{\bm{\Pi}}\x\right\rangle=-i\sum_{\mb{q}\sigma\tau}\frac{A\uq^2}{\lambda\uqt\sqrt{\lambda\uqt}}\bm{\epsilon}\uqs\bigg(\bm{\epsilon}\uqt\cdot\Big[H_{\mb{q}\sigma\tau}\x\bm{f}\uq-H^*_{\mb{q}\sigma\tau}\x\bm{f}^\dagger\uq\Big]\bigg),
\end{equation}
where $\sqrt{2}\bm{\epsilon}\uqt=\tau\bm{\epsilon}_{\mb{q}1}+\bm{\epsilon}_{\mb{q}2}$ and 
\begin{align}
H_{\mb{q}1\tau}\x&=w\uqt\rme^{i\mb{q}\cdot\mb{x}}+y\uqt\rme^{-i\mb{q}\cdot\mb{x}},\\
H_{\mb{q}2\tau}\x&=x\uqt\rme^{i\mb{q}\cdot\mb{x}}+z\uqt\rme^{-i\mb{q}\cdot\mb{x}}.
\end{align}

The Coulomb-gauge has $\mb{g}_\text{T}^C(\mb{x},\mb{x}')=\mb{0}$ and the dipole-gauge is the long wavelength limit of multipolar-gauge, which is defined by $g_{\text{T}i}^m(\mb{x},\mb{x}')=-\intl x'_j\delta_{ij}^\text{T}(\mb{x}-\lambda\mb{x}')$. Therefore, the average transverse electric field calculated in these gauges is
\begin{align}
\left\langle\hat{ E}_{\text{T}i}\x\right\rangle^C&=-\left\langle\hat{\Pi}_{i}\x\right\rangle^C,\\
\left\langle \hat{E}_{\text{T}i}\x\right\rangle^d&=-\left\langle\hat{\Pi}_{i}\x\right\rangle^d-\left\langle \hat{D}_j\right\rangle^d_m\delta_{ij}^\text{T}\x,\label{eq:ETD}
\end{align}
where $\hat{D}_j=\intxp x'_j\hat{\rho}\xp$ is the total dipole operator. Below we prove that in both gauges $\langle\hat{\mb{E}}_\text{T}\x\rangle=\mb{0}$.

\subsection{Dipole-gauge}
The dipole-gauge has $\lambda\uqt=1$, $\hat{\mb{M}}^p\uq=\hat{\mb{0}}$ and $\hat{P}_{\text{T}\mb{q}i}=\hat{D}_j\delta_{\mb{q}ij}^\text{T}$ where $\delta_{\mb{q}ij}^\text{T}=(1/V)\intx\delta_{ij}^\text{T}\x\rme^{-i\mb{q}\cdot\mb{x}}$. The first property means that $w\uqt=-\tau/\sqrt{2}$, $x\uqt=-1/\sqrt{2}$ and $y\uqt=z\uqt=0$. The final two properties mean that $\hat{f}_{\mb{q}i}=iV\nu\uq\hat{D}_j\delta_{\mb{q}ij}^\text{T}$. Substituting these, along with $A\uq^2=1/(2\nu\uq V)$, into Eq.~\eqref{eq:PiExp} we find
\begin{equation}
\left\langle\hat{\bm{\Pi}}\x\right\rangle^D=-\frac{1}{2}\sum_{\mb{q}\tau}\bm{\epsilon}\uqt\bigg(\bm{\epsilon}\uqt\cdot\Big[\rme^{i\mb{q}\cdot\mb{x}}\mb{X}\uq^D+\rme^{-i\mb{q}\cdot\mb{x}}[\mb{X}\uq^D]^\dagger\Big]\bigg),
\end{equation}
where we have defined the dipole-gauge coupling vector,
\begin{equation}
X^D_{\mb{q}i}=\left\langle \hat{D}_j\right\rangle^D_m\delta_{\mb{q}ij}^\text{T}.
\end{equation}
Hence,
\begin{align}
\left\langle\hat{\Pi}_{i}\x\right\rangle^D&=-\frac{1}{2}\sum_{\mb{q}\tau}\epsilon_{\mb{q}\tau i}\bigg(\bm{\epsilon}\uqt\cdot\left\langle\hat{\mb{D}}\right\rangle^D_m\rme^{i\mb{q}\cdot\mb{x}}+\text{c.c.}\bigg)\\
&=-\sum_{\mb{q}\tau}\epsilon_{\mb{q}\tau i}\left(\bm{\epsilon}\uqt\cdot\left\langle\hat{\mb{D}}\right\rangle^D_m\right)\rme^{i\mb{q}\cdot\mb{x}}\\
&=-\left\langle \hat{D}_j\right\rangle^D_m\delta^\text{T}_{ij}\x,
\end{align}
where to arrive at the second line we have relabelled $\mb{q}\to-\mb{q}$ in the conjugate term and used that $\langle \hat{D}_j\rangle^D_m\in\Re$. Substituting this into Eq.~\eqref{eq:ETD} gives $\langle \hat{\mb{E}}_\text{T}\rangle^D=\mb{0}$.

\subsection{Coulomb-gauge}
The Coulomb-gauge has $\lambda\uqt=\sqrt{1+[(e^2 N )/(m V\nu\uq^2)]}\equiv\lambda\uq$, $\hat{\mb{P}}_\text{T}=\hat{\mb{0}}$ and $\hat{\mb{M}}^p\uq=\hat{\mb{M}}^p_{0\mb{q}}=i(\check{\mb{q}}\times\hat{\mb{j}}^p\uq)/\nu\uq$ where,
\begin{equation}
\hat{\mb{j}}^p\uq=-\frac{e}{2mV}\sum_\mu\{\hat{\mb{p}}_\mu,\rme^{-i\mb{q}\cdot\hat{\mb{r}}_\mu}\}.
\end{equation}
The first of these properties gives
\begin{equation}
H_{\mb{q}2\tau}\x=-\frac{1}{2\sqrt{2\lambda\uq}}\Big(\left[\lambda\uq+1\right]\rme^{i\mb{q}\cdot\mb{x}}+\left[\lambda\uq-1\right]\rme^{-i\mb{q}\cdot\mb{x}}\Big),
\end{equation}
and $H_{\mb{q}1\tau}\x=\tau H_{\mb{q}2\tau}\x$. The final two properties lead to $\hat{\bm{f}}\uq=-V[\check{\mb{q}}\times(\check{\mb{q}}\times\hat{\mb{j}}^p\uq)]$. Substituting these, along with the identity
\begin{equation}
\bm{\epsilon}\uqs\cdot\left[\check{\mb{q}}\times(\check{\mb{q}}\times\hat{\mb{j}}^p\uq)\right]=-\bm{\epsilon}\uqs\times\hat{\mb{j}}^p\uq,
\end{equation}
into Eq.~\eqref{eq:PiExp} we find that
\begin{equation}\label{eq:ETC}
\left\langle\hat{\bm{\Pi}}\x\right\rangle^C=-\frac{1}{2}\sum_{\mb{q}\tau}\bm{\epsilon}\uqt\bigg(\bm{\epsilon}\uqt\cdot\Big[\rme^{i\mb{q}\cdot\mb{x}}\mb{X}\uq^C-\rme^{-i\mb{q}\cdot\mb{x}}[\mb{X}\uq^C]^\dagger\Big]\bigg),
\end{equation}
where the Coulomb-gauge coupling vector is
\begin{equation}
\mb{X}^C\uq=\frac{-i}{\nu\uq\lambda\uq^2}\left( \left[\lambda\uq+1\right]\left\langle \hat{\mb{j}}^p\uq\right\rangle^C_m+\left[\lambda\uq-1\right]\left\langle \hat{\mb{j}}^{p\dagger}\uq\right\rangle^C_m\right).
\end{equation}
To coincide with the assumptions made in the dipole-gauge we must now take the long wavelength limit, $\exp[\pm i\mb{q}\cdot\hat{\mb{r}}_\mu]\to1$. In this limit,
\begin{equation}
    \lim_{\exp[\pm i\mb{q}\cdot\hat{\mb{r}}_\mu]\to1}\hat{\mb{j}}^p\uq=-\frac{e}{2mV}\sum_\mu\hat{\mb{p}}_\mu.
\end{equation}
Hence, one can write $\lim\{\exp[\pm i\mb{q}\cdot\hat{\mb{r}}_\mu]\to1\}\hat{j}^p_{\mb{q}i}=-i[\hat{d}_i,\hat{H}_m]$, where $\hat{d}_i=-e\sum_\mu\hat{d}_{\mu i}$ is the total electron dipole operator and $\hat{H}_m$ is the matter Hamiltonian defined in the main text above Eq.~\eqref{eq:HE}. This, along with the fact that in the thermodynamics limit $\ket{\psi_m}$ is an eigenstate of the $\hat{H}_m$ with eigenvalue $\epsilon_m$, we can write that \cite{andolina2019cavity}
\begin{equation}
\lim_{\exp[\pm i\mb{q}\cdot\hat{\mb{r}}_\mu]\to1}\left\langle \hat{j}^{p}_{\mb{q}i}\right\rangle_m=-i\braket{\psi_m\left|\left[\hat{d}_i,\hat{H}_m\right]\right|\psi_m}=-i\left[\epsilon_m-\epsilon_m\right]\braket{\psi_m|\hat{d}_i|\psi_m}=0.
\end{equation}
Therefore, $\langle\mb{E}_\text{T}\rangle^C=\mb{0}=\langle\mb{E}_\text{T}\rangle^D$, verifying the gauge-invariance of the prediction.
	
\end{document}